\documentclass[12pt]{article}
\usepackage{amsfonts,graphicx,amssymb,amsmath,amsthm,epsfig}
\usepackage{float}
\allowdisplaybreaks
\begin{document}
\title{\textbf{Charged Gravastar Solutions in the Finch-Skea Framework with
$f(\mathbb{Q})$ Gravity}}
\author{M. Sharif \thanks {msharif.math@pu.edu.pk}~~and
Madiha Ajmal \thanks {madihaajmal222@gmail.com} \\
Department of Mathematics and Statistics, The University of Lahore,\\
1-KM Defence Road Lahore-54000, Pakistan.}

\date{}
\maketitle
\begin{abstract}
In this study, we investigate the features of a charged gravastar
within the framework of $f(\mathbb{Q})$ gravity ($\mathbb{Q}$
represents non-metricity) using the Finch-Skea metric. This metric
is applied to both the interior and shell regions of the charged
gravastar and the field equations are derived accordingly. For the
exterior regions, we consider various black holes, i.e.,
Reissner-Nordstr$\ddot{o}$m, Bardeen and Hayward regular black
holes. The interior and exterior layers are matched using the Israel
junction conditions, which help to determine the surface energy
density and surface pressure for these black holes. We examine some
physical properties such as proper length, entropy, energy and the
equation of state parameter. The stability of the developed
gravastar model is discussed through  the effective potential, the
causality condition and adiabatic index. We conclude that the
compact gravastar structure could be a viable alternative to black
holes within this framework.
\end{abstract}
\textbf{Keywords}: $f(\mathbb{Q})$ gravity; Israel formalism;
Gravastars.\\
\textbf{PACS}: 04.50.kd; 04.70.Bw; 04.40.Dg.

\section{Introduction}

In the early 20th century, Albert Einstein revolutionized our
understanding of the universe with the introduction of general
theory of relativity (GR). In modern cosmology, observations have
revealed that the universe is undergoing an accelerated expansion,
driven by a mysterious force known as dark energy (DE) \cite{aa}.
Although the cosmological constant in GR can explain the expanding
cosmos \cite{bb}, it encounters challenges such as cosmic
coincidence \cite{cc} and fine-tuning problems \cite{dd}. In this
regard, GR may not be the most suitable model for explaining gravity
on large scales. Our understanding of cosmic acceleration is still
limited, hence the ongoing research into modified theories of
gravity is crucial. These studies offer promising alternatives to GR
and could address some of the existing challenges. Over the past
decades, extensive research has focused on modified theories of
gravity to deepen our knowledge of the universe structure \cite{ee}.

In GR, the structure of spacetime is typically characterized through
Riemannian geometry, utilizing the Levi-Civita connection to
describe the motion of objects within this framework \cite{1}. This
connection assumes that only curvature exists, while two other
geometric properties, non-metricity $(\mathbb{Q})$ and torsion
$(\mathbb{T})$ are considered to be absent. However, there are
different ways to define connections on a manifold, which can lead
to different but equivalent descriptions of gravity. By allowing
non-metricity and torsion to exist, we can develop alternative
theories of gravity. If we relax the condition on torsion but keep
curvature and non-metricity zero, we get the teleparallel equivalent
of GR \cite{2}. Alternatively, a flat spacetime that incorporates
non-metricity without torsion leads to the symmetric teleparallel
gravity of GR \cite{3} and its extension is known as $f(\mathbb{Q})$
gravity, where gravitational interactions are defined through
$\mathbb{Q}$ \cite{6}. The motivation for choosing $f(\mathbb{Q})$
gravity \cite{7} lies in its ability to address challenges in
cosmology and astrophysics, particularly in explaining the
accelerated expansion of the cosmos without requiring DE. This
theory modifies the GR framework by allowing the action to be a
general function of $\mathbb{Q}$, which offers new avenues for
exploring gravitational phenomena.

Researchers are highly motivated to delve into $f(\mathbb{Q})$
theory. Harko et al. \cite{g1} studied this theory using power-law
and exponential forms to understand how matter interacts with this
theory. Lazkoz et al. \cite{g2} conducted an observational analysis
of different $f(\mathbb{Q})$ models using redshift-based
parametrization to assess their viability as alternatives to the
$\Lambda$CDM model for explaining the universe late-time
acceleration. Mandal et al. \cite{g3} explored this gravity through
energy conditions to identify viable models that are compatible with
the rapid expansion of the cosmos. Lin and Zhai \cite{g4}
investigated this gravity to explore the effects of $f(\mathbb{Q})$
on compact stars, revealing that a negative modification supports
more stellar mass while a positive modification reduces it. Mandal
and Sahoo \cite{g6} calculated the equation of state (EoS) as well
as Hubble parameters for Pantheon samples and showed that the
$f(\mathbb{Q})$ model behaves differently from the standard cold
dark matter. Lymperis \cite{g7} examined the cosmological effects
through the effective DE sector in this theory. Koussour et al.
\cite{g8} studied how cosmic parameters behave in this gravity. de
Araujo and Fortes \cite {g9} analyzed this gravity models by
applying them to polytropic stars and examined non-metricity
behavior both inside and outside the stars. In recent studies
\cite{g10}, we have developed the generalized ghost DE and
generalized ghost pilgrim DE models within the same gravity in an
interacting scheme. We have also examined the pilgrim and
generalized ghost pilgrim DE models under a non-interacting scenario
\cite{g11}.

The study of stellar structures and gravitational collapse has been
a key focus in astrophysics since the development of GR.
Gravitational collapse leads to the formation of compact objects
such as neutron stars, white dwarfs and black holes, with the final
outcome depending on the initial mass of the star. Both theoretical
and observational perspectives have extensively explored these
remnants. One of the most fascinating and complex topics in modern
astrophysics is the study of compact objects, particularly black
holes. Black holes result from the complete gravitational collapse
of massive stars, a phenomenon explained by general relativity (GR).
They contain singularities at their centers and are surrounded by an
event horizon, a boundary that nothing, not even light, can escape.
This event horizon acts as a one-way barrier, separating the
interior of the black hole from the outside universe. Strong
evidence for the existence of supermassive black holes has been
found, especially in the case of Sagittarius A*, located at the
center of our Milky Way galaxy, which has reinforced our
understanding of black holes.

In the light of challenges presented by black holes, Mazur and
Mottola \cite{1a} proposed the concept of gravitational vacuum stars
or gravastars, as an alternative. Unlike black holes, gravastars do
not have event horizons or singularities. Instead, they are
envisioned as extremely dense objects with a core of negative
pressure, surrounded by a shell of matter, possibly exotic, which
prevents singularity formation \cite{f09}. This model offers a
potential solution to some of the key issues associated with black
holes, such as the information loss paradox and the problematic
nature of singularities. While the Schwarzschild solution in 1916
laid the foundation for understanding black holes, it has
limitations, particularly regarding the unresolved questions
surrounding singularities and event horizons \cite{f10}. Gravastars,
with their unique structure, may provide a more stable endpoint for
stellar evolution, offering an alternative to black holes. Although
direct observational evidence for gravastars is still lacking,
studying them could address some conceptual challenges in
understanding black holes and stellar collapse, which has long
intrigued astrophysicists \cite{4ee}.

There has been a significant research on gravastars, primarily
within the framework of GR \cite{f11}-\cite{f44}. Recent
observational evidence such as the accelerated expansion of the
universe and the existence of dark matter have presented theoretical
challenges to this framework \cite{f55}. While GR has been pivotal
in uncovering many mysteries of the universe, it falls short in
explaining these newer phenomena. Sakai et al. \cite{2abb} explored
the shadows of gravastars as a way to identify them. Kubo and Sakai
\cite{3abb} proposed that gravitational lensing, a phenomenon where
light bends around massive objects, could help to find gravastars,
since black holes do not show the same microlensing effects with
maximum brightness. Das et al. \cite{mn} proposed a unique stellar
model based on the gravastar concept under $f(\mathbb{R},T)$
gravity, where $\mathbb{R}$ and $T$ represent the Ricci scalar and
trace of the energy-momentum tensor (EMT), respectively. This
gravastar consists of three regions described by de Sitter spacetime
and provides exact singularity-free solutions with valid physical
properties within alternative gravity.

Shamir and Ahmad \cite{m2} explored the structure of gravastars in
$f(\mathbb{G}, T)$ gravity (where $\mathbb{G}$ is the Gauss-Bonnet
term) and examined various physical aspects, indicating that
gravastars could provide stable non-singular solutions without an
event horizon. Bhatti et al. \cite{n1} explored a singularity-free
gravastar model in $f(\mathbb{R},\mathbb{G})$ gravity and discussed
physical properties, highlighting the role of this gravity in the
model's stability. Sengupta et al. \cite{df3} constructed a
gravastar model within the Randall-Sundrum braneworld gravity
framework, calculating solutions and physical parameters including
surface redshift to assess its stability. Bhatti et al. \cite{f6}
analyzed a gravastar structure in $f(\mathbb{G},T^{2})$ gravity and
explored the properties of the inner and central layers to deduce
parameters of self-gravitating structure. Sinha and Singh \cite{f7}
studied gravastars in $f(\mathbb{R}, L_m,T)$ theory, focusing on the
negative energy density in the interior, the repulsive force on the
thin-shell and the exterior described by the Schwarzschild-de Sitter
solution. Teruel et al. \cite{df2} presented the first discovery of
gravastar configurations in modified gravity theory, identifying two
physically feasible models with regular interiors and specific shell
properties. Sharif et al. \cite{f8} introduced a new solution for
the gravastar model in non-conservative Rastall gravity, derived
from Tolman IV spacetime.

Charge is hard to see directly in compact objects like stars or
black holes but it can build up in extreme space environments.
Strong magnetic fields around objects like pulsars and black holes
can lead to charged areas. This happens through processes like
pulling in charged particles and interacting with ionized
(electrically charged) gas. Incorporating charge into the gravastar
model allows us to account for these realistic conditions, exploring
how the interplay between electromagnetic forces and gravity
influences the stability of such objects. This approach broadens the
theoretical scope of gravastar models, aligning them more closely
with potential observational realities where charge may play a role.

The study of charge in different modified theories has attracted
significant interest because of its intriguing characteristics.
$\ddot{O}$vg$\ddot{u}$n et al. \cite{1kk} investigated charged
thin-shell gravastars within non-commutative geometry. Sharif and
Waseem \cite{m1} analyzed how charge affects a gravastar within
$f(\mathbb{R},T)$ gravity and found non-singular, physically
consistent solutions with various properties. Bhar and Rej \cite{f3}
focused on the configuration and stability of a charged gravastar in
$f(\mathbb{R},T)$ gravity. Barzegar et al. \cite{f5} investigated
three-dimensional anti-de Sitter gravastars in the context of
gravity's rainbow theory. Their findings revealed that the physical
parameters for charged and uncharged states depend on rainbow
functions and the electric field. Ilyas et al. \cite{f51}
investigated charged compact stars within the extended gravitational
theory $f(\mathbb{R},\mathbb{G},T)$, proposing models to analyze
physical properties of relativistic objects. Bhar \cite{f53}
investigated self-gravitating structures of an anisotropic fluid in
the torsion-dependent extended theory of gravity while determining
the maximum mass and radius for four compact stars. Bhattacharjee
and Chattopadhyay \cite{f54} extended the gravitational
Bose-Einstein condensate star concept to a charged gravastar system,
resulting in lower entropy than that of black holes and stability
without the information paradox. The same authors \cite{f555}
analyzed the impact of charge on the formation of a gravastar in
Rastall gravity, with stability ensured through gravitational
surface redshift and entropy maximization.

Recently, there has been a great interest about the study of
gravastars in $f(\mathbb{Q})$ gravity. The gravastar model has been
examined under various gravitational theories, including
$f(\mathbb{Q})$ gravity \cite{5ee}. Pradhan et al. \cite{6aa}
presented a gravastar model using the Mazur-Mottola method in this
theory, focusing on an isotropic matter distribution. Mohanty and
Sahoo \cite{8aa} investigated gravastar characteristics using the
Krori-Barua metric and examined its physical properties. Javed et
al. \cite{7bb} investigated how charge affects gravastars in
$f(\mathbb{Q})$ gravity. Mohanty et al. \cite{6bb} studied charged
gravastar and discussed its energy density, entropy and EoS. They
also discussed how future radio telescopes could detect the
gravastar shadow, distinguishing it from a black hole event horizon.

The Finch-Skea metric has been employed by various researchers to
study gravastars, wormholes and strange stars in GR and other
gravitational theories. Sharma and Ratanpal \cite{gh1} presented a
class of solutions for static spherically symmetric anisotropic
stars using the Finch-Skea ansatz, which led to a quadratic equation
of state with physically justified parameter bounds. Bhar et al.
\cite{hh4} introduced a new interior solution for a (2 +
1)-dimensional anisotropic star in Finch-Skea spacetime related to
the Banados, Teitelboim and Zanelli black hole. Banerjee et al.
\cite{hh5} provided exact solutions to the Einstein field equations
using the Finch-Skea ansatz, confirming alignment with the observed
maximum mass for pulsars. Dey and Paul \cite{hh2} proposed
relativistic solutions for compact objects using Finch-Skea
geometry, showing that stars are isotropic and uncharged in 4D.
These solutions help to construct realistic stellar models. Maurya
et al. \cite{hh3} developed an anisotropic stellar model using the
gravitational decoupling technique with the Finch-Skea metric.
Majeed et al. \cite{f65} introduced a gravastar model in
$f(\mathbb{R},T)$ gravity using the Finch-Skea function, resulting
in a thermodynamically stable collapse free from black hole theory
issues.

Sharif and Naz \cite{1gg} investigated the formation of a gravastar
in $f(\mathbb{R},T^{2})$ theory using the Finch-Skea metric.
Dayanandan et al. \cite{f2bb} presented a deformed Finch-Skea
anisotropic solution using Class I spacetime, applying gravitational
decoupling to model compact objects. Sharif and Manzoor \cite{1hh}
examined the viability and stability of anisotropic compact stars in
the same theory using Finch-Skea symmetry. Gul et al. \cite{f2aa}
investigated the viability and stability of anisotropic compact
stars in $f(\mathbb{Q})$ theory, analyzed their physical properties
using Finch-Skea solutions, and confirmed their stability through
various criteria. Mustafa et al. \cite{3gg} examined the dynamics of
a spherically symmetric, anisotropic compact star in $f(\mathbb{Q})$
gravity using the Karmarkar condition and Finch-Skea structure.
Karmakar et al. \cite{2gg} introduced a model in 5D
Einstein-Gauss-Bonnet gravity with the Finch-Skea ansatz, applied it
to EXO 1785-248 and confirmed that it met all physical criteria.

Rej et al. \cite{f61} developed a relativistic, anisotropic DE
stellar model utilizing the complexity factor formalism and
Finch-Skea ansatz. Shahzad et al. \cite{f62} introduced a new
solution in Rastall theory with a quintessence field, employing the
Finch-Skea ansatz to analyze isotropic matter in compact stars. Das
et al. \cite{f64} investigated a relativistic model of anisotropic
compact objects, focusing on the Finch-Skea metric and its
application to pulsar PSR J0348+0432. Gul et al. \cite{f2cc} studied
the viability and stability of anisotropic stellar objects in
modified symmetric teleparallel gravity, utilizing the Finch-Skea
solution.

In this article, we explore the charged gravastar model within the
$f(\mathbb{Q})$ framework using the Finch-Skea metric. The format of
the paper is given as follows. In section \textbf{2}, we briefly
discuss $f(\mathbb{Q})$ gravity with Maxwell equations. We analyze
the corresponding field equations for a spherically symmetric
spacetime with the Finch-Skea metric potentials. Section \textbf{3}
focuses on the structure of a charged gravastar. In section
\textbf{4}, we explore boundary and junction conditions. Section
\textbf{5} discusses physical characteristics of the developed
gravastar model. We also analyze stability of the model. Section
\textbf{6} presents the conclusion of our findings.

\section{\textbf{Structure of $f(\mathbb{Q})$ Gravity}}

In the $f(\mathbb{Q})$ gravity theory, the non-metricity is
described as \cite{40a}
\begin{equation}\label{1a}
\mathbb{Q}_{\lambda\iota\vartheta}=-g_{\iota\vartheta,\lambda}+g_{\vartheta\sigma}
\hat{\Gamma}^{\sigma}_{\iota\lambda}
+g_{\sigma\iota}\hat{\Gamma}^{\sigma}_{\vartheta\lambda},
\end{equation}
where the metric tensor is represented by $g_{\iota\vartheta}$ and
the affine connection is denoted as
$\hat{\Gamma}^{\lambda}_{\iota\vartheta}$. The affine connection can
be divided into three distinct components \cite{41b}
\begin{equation}\label{2a}
\hat{\Gamma}^{\lambda}_{\iota\vartheta}={\Gamma}^{\lambda}_{\iota\vartheta}
+\mathbb{C}^{\lambda}_{\;\iota\vartheta}+\mathbb{L}^{\lambda}_{\;\iota\vartheta}.
\end{equation}
Here, the Levi-Civita connection is given by
\begin{equation}\label{3a}
\Gamma^{\lambda}_{\iota\vartheta}=\frac{1}{2}g^{\lambda\sigma}
(g_{\sigma\vartheta,\iota}+g_{\sigma\iota,\vartheta}-g_{\iota\vartheta,\sigma}),
\end{equation}
the contortion tensor is expressed as
\begin{equation}\label{4a}
\mathbb{C}^{\lambda}_{\;\iota\vartheta}=\hat{\Gamma}^{\lambda}_{[\iota\vartheta]}
+g^{\lambda\sigma}g_{\iota\kappa}\hat{\Gamma}^{\kappa}_{[\vartheta\sigma]}
+g^{\lambda\sigma}g_{\vartheta\kappa}\hat{\Gamma}^{\kappa}_{[\iota\sigma]},
\end{equation}
and the disformation tensor is described as
\begin{equation}\label{5a}
\mathbb{L}^{\lambda}_{\;\iota\vartheta}=\frac{1}{2}g^{\lambda\sigma}(\mathbb{Q}_{\vartheta\iota\sigma}
+\mathbb{Q}_{\iota\vartheta\sigma}-\mathbb{Q}_{\sigma\iota\vartheta}).
\end{equation}
Next, the superpotential is defined as
\begin{equation}\label{6a}
\mathbb{P}^{\lambda}_{\;\iota\vartheta}=-\frac{1}{2}\mathbb{L}^{\lambda}_{\;\iota\vartheta}
+\frac{1}{4}(\mathbb{Q}^{\lambda}-\tilde{\mathbb{Q}}^{\lambda})g_{\iota\vartheta}-
\frac{1}{4} \delta ^{\lambda}\;_{({\iota}}\mathbb{Q}_{\vartheta)},
\end{equation}
where
\begin{equation}\label{7a}
\mathbb{Q}_{\lambda}=\mathbb{Q}^{~\iota}_{\lambda~\iota},\quad
\tilde{\mathbb{Q}}_{\lambda}=\mathbb{Q}^{\iota}_{~\lambda\iota}.
\end{equation}
The non-metricity scalar can be described as \cite{32-a}
\begin{equation}\label{8a}
\mathbb{Q}=-\mathbb{Q}_{\lambda\iota\vartheta}\mathbb{P}^{\lambda\iota\vartheta}
=-\frac{1}{4}(-\mathbb{Q}^{\lambda\vartheta\upsilon}\mathbb{Q}_{\lambda\vartheta\upsilon}
+2\mathbb{Q}^{\lambda\vartheta\upsilon}\mathbb{Q}_{\upsilon\lambda\vartheta}-2\mathbb{Q}^{\upsilon}\tilde{\mathbb{Q}}_{\upsilon}+\mathbb{Q}^{\upsilon}\mathbb{Q}_{\upsilon}).
\end{equation}

The action for $f(\mathbb{Q})$ gravity, including both the matter
Lagrangian $L_{m}$ and the electromagnetic field Lagrangian
$L_{\mathbf{e}}$, becomes \cite{6}
\begin{equation}\label{9a}
S=\int\frac{1}{2}f(\mathbb{Q})
\sqrt{-g}d^{4}x+\int(L_{m}+L_{\mathbf{e}}) \sqrt{-g}d^{4}x,
\end{equation}
where
\begin{equation}\label{10a}
L_{\mathbf{e}}=-\frac{1}{16\pi} F_{\mu\nu} F^{\mu\nu}.
\end{equation}
The Maxwell field tensor $ F_{\mu\nu}$ is defined as
$\varphi_{\mu,\nu} - \varphi_{\nu,\mu} $, with $\varphi_{\mu}$
representing the four-potential. The field equations for
$f(\mathbb{Q})$ gravity are expressed as
\begin{equation}\label{11a}
\frac{-2}{\sqrt{-g}}\nabla_{\lambda}(f_{\mathbb{Q}}\sqrt{-g}
\mathbb{P}^{\lambda}_{\;\iota\vartheta})-\frac{1}{2}f
g_{\iota\vartheta}-f_{\mathbb{Q}}
(P_{\iota\varpi\varsigma}\mathbb{Q}_{\vartheta}^{~\varpi\varsigma}-2\mathbb{Q}^{\varpi\varsigma}_{~~~\iota}
P_{\varpi\varsigma\vartheta})=T_{\iota\vartheta}+\mathbb{E}_{\iota\vartheta},
\end{equation}
where the EMT is
\begin{equation}\label{29}
T_{\iota\vartheta} \equiv \frac{-2}{\sqrt{-g}} \frac{\delta
(\sqrt{-g} L_{M})}{\delta g^{\iota\vartheta}},
\end{equation}
and $f_{\mathbb{Q}}$ is the derivative with respect to $\mathbb{Q}$.
The EMT for the electromagnetic field describes how the
electromagnetic field affects the curvature of spacetime. This is
given by
\begin{equation}\label{12a}
\mathbb{E}_{\iota\vartheta} = \frac{1}{4\pi} \bigg( F^{\nu}_\iota
F_{\vartheta\nu} - \frac{1}{4} g_{\iota\vartheta} F_{\mu\nu}
F^{\mu\nu} \bigg).
\end{equation}
Maxwell equations describe how electric and magnetic fields work and
interact with each other. In curved spacetime, these equations are
given as
\begin{equation}\label{13a}
(\sqrt{-g} F_{\iota\vartheta})_{;\vartheta} = 4\pi J_\iota
\sqrt{-g}, \quad F_{[\iota\vartheta;\delta]} = 0,
\end{equation}
where $J_\iota = \sigma u_\iota$ is the electric four-current, with
$\sigma$ indicating the charge density. The electric field strength
is described by \cite{2bb}
\begin{equation}\label{14a}
\mathbb{E}(r) = \frac{e^{\frac{\alpha + \beta}{2}}}{r^2}
\mathcal{E}(r),
\end{equation}
here $\mathcal{E}(r)$ represents the total charge contained within a
sphere of radius $r$ and is determined as follows
\begin{equation}\label{15a}
\mathcal{E}(r) = 4\pi \int_0^r \sigma r^2 e^\beta \, dr.
\end{equation}

The spherically symmetric interior metric in the standard form
follows
\begin{equation}\label{1}
ds_-^{2}=-e^{\alpha(r)}dt^{2}+e^{\beta(r)}dr^{2}+r^{2}(d\theta^{2}+\sin^{2}\theta
d\phi^{2}),
\end{equation}
where $``-"$ indicates the interior metric. The EMT for a perfect
fluid can be expressed as
\begin{equation}\label{9aa}
T_{\iota\vartheta}=(\varrho+p)u_{\iota}u_{\vartheta}+pg_{\iota\vartheta},
\end{equation}
where $\varrho$ represents the energy density of the fluid, while
$p$ is the pressure and the four-velocity components of the fluid
are denoted as $u_{\iota}$. In $f(\mathbb{Q})$ gravity, the
Einstein-Maxwell field equations are expressed as follows
\begin{eqnarray}\label{2}
\varrho+2 \pi
\mathcal{E}^{2}&=&\frac{f}{2}+\frac{2}{r}f_{\mathbb{Q}\mathbb{Q}}e^{-\beta}\mathbb{Q}^{\prime}-
f_{\mathbb{Q}}\bigg[
\mathbb{Q}+\frac{e^{-\beta}}{r}(\alpha^{\prime}+\beta^{\prime})+\frac{1}{r^{2}}\bigg],
\\\label{3} p-2 \pi
\mathcal{E}^{2}&=&-\frac{f}{2}+ f_{\mathbb{Q}}\bigg(
\mathbb{Q}+\frac{1}{r^{2}}\bigg),
\\\nonumber p+2 \pi
\mathcal{E}^{2}&=& f_{\mathbb{Q}}
\bigg(\frac{\mathbb{Q}}{2}-e^{-\beta} \bigg(\frac{\alpha^{\prime
\prime}}{2}+\bigg(\frac{\alpha^{\prime}}{4}+\frac{1}{2 r}\bigg)
\big(\alpha^{\prime}-\beta^{\prime}\big)\bigg)\bigg)\\\label{4}&-&f_{\mathbb{Q}\mathbb{Q}}
e^{-\beta} \mathbb{Q}^{\prime}\bigg(\frac{\alpha
^{\prime}}{2}+\frac{1}{r}\bigg)-\frac{f}{2}.
\end{eqnarray}
The non-metricity scalar is given as \cite{8aa}
\begin{equation}\label{5}
\mathbb{Q}=-\frac{\big(\alpha^{\prime}(r)+\frac{1}{r}\big) \big(2
e^{-\beta (r)}\big)}{r},
\end{equation}
where prime represents derivative with respect to $r$. The linear
model is the well-known initial model of $f(\mathbb{Q})$ theory
\cite{4aa}
\begin{equation}\label{6}
f(\mathbb{Q})=\zeta +\mathbb{Q} \psi,
\end{equation}
with $\psi$ is a non-zero constant and $\zeta$ as arbitrary
constant. Here $f_{\mathbb{Q}}=\psi$ and
$f_{\mathbb{Q}\mathbb{Q}}=0$. To recover GR from this modified
theory, we consider the linear form of the function,
$f(\mathbb{Q})=\zeta +\mathbb{Q} \psi$, where $\psi$ and $\zeta$ are
constants. Specifically, when $\psi = -1$ and $\zeta = 0$, the
theory reduces to the Einstein-Hilbert action, which underpins GR
\cite{8aa}. Under this assumption, the field equations derived from
$f(\mathbb{Q})$ gravity become identical to those of GR, where the
metric alone describes the curvature of spacetime and non-metricity
vanishes. Substituting Eqs.\eqref{5} and \eqref{6} into \eqref{2}
through \eqref{4}, we obtain
\begin{eqnarray}\label{7}
\varrho&=&\frac{e^{-\beta} \bigg(e^{\beta} \big(r^2 (\zeta -4 \pi
\mathcal{E}^{2})-2 \psi \big)-2 r \psi  \beta^{\prime}+2 \psi
\bigg)}{2 r^2},\\\label{8} p&=&\frac{e^{-\beta (r)} \bigg(e^{\beta}
\big(r^2 (4 \pi  \mathcal{E}^{2}-\zeta )+2 \psi \big)-2 \psi\big(r
\alpha^{\prime}+1\big)\bigg)}{2 r^2},\\
\label{9} p&=&\frac{e^{-\beta} \bigg(-2 r (4 \pi
\mathcal{E}^{2}+\zeta ) e^{\beta}-2 r \psi
\alpha^{\prime\prime}-\psi  \big(r \alpha^{\prime}+2\big)
\big(\alpha^{\prime}-\beta^{\prime}\big)\bigg)}{4 r}.
\end{eqnarray}

We select the Finch-Skea metric in this study for its unique
mathematical properties, making it well-suited for modeling compact
objects like gravastars. Its primary advantage is that it provides
regular and physically realistic solutions for the interior of
spherically symmetric objects, avoiding singularities. This ensures
well-behaved spacetime under extreme gravitational conditions.
Additionally, its simplicity allows for easier analytical handling,
which is crucial when dealing with non-linear field equations in
systems like charged gravastars within $f(\mathbb{Q})$ gravity.

The Finch-Skea solutions, which are regarded important for obtaining
feasible interior solutions, are given as follows \cite{23aa}
\begin{equation}\label{10}
e^{\alpha(r)}=\left(\xi +\frac{1}{2} r \varphi  \sqrt{r^2
\chi}\right)^2,\quad e^{\beta(r)}=r^2 \chi +1,
\end{equation}
here $\xi$, $\varphi$ and $\chi$ are unknown non-zero constants
whose values can be determined by applying matching conditions.
Inserting Eq.\eqref{10} into \eqref{7} through \eqref{9}, we have
\begin{eqnarray}\label{11}
\varrho&=&\frac{1}{2} \bigg(\zeta-4 \pi\mathcal{E}^{2} -\frac{2 \chi
\psi \big(r^2 \chi +3\big)}{\big(r^2 \chi +1\big)^2}\bigg),
\\\label{12} p&=&-\frac{-4 \pi \mathcal{E}^{2} r \big(r^2 \chi +1\big)
+\zeta  r^3 \chi +\frac{8 \varphi  \psi  \sqrt{r^2 \chi }}{2 \xi +r
\varphi  \sqrt{r^2 \chi }}+r (\zeta -2 \chi  \psi )}{2 \big(r^3 \chi
+r\big)},
\\\nonumber p&=&\frac{-1}{2 \big(r^2 \chi
+1\big)^2 \big(r^3 \varphi  \chi +2 \xi  \sqrt{r^2 \chi
}\big)}\bigg(4 \pi \mathcal{E}^{2}\big(r^3 \varphi \chi +2 \xi
\sqrt{r^2 \chi }\big)
\\\nonumber&\times&\big(r^2 \chi +1\big)^2+r^3 \varphi  \chi (\zeta +2 \chi  \psi )
+2 \xi\sqrt{r^2 \chi } (\zeta -2 \chi  \psi )+4
\\\label{13}&\times& \zeta\xi\big(r^2 \chi \big)^{\frac{3}{2}}+2 \zeta
r^5 \varphi  \chi ^2+2 \zeta \xi \big(r^2 \chi
\big)^{\frac{5}{2}}+\zeta r^7 \varphi \chi ^3+8 r \varphi \chi  \psi
\bigg).
\end{eqnarray}
Subtracting Eqs.\eqref{12} and \eqref{13}, it follows that
\begin{eqnarray}\label{16}
\mathcal{E}^{2}&=&-\frac{r \chi  \psi  \bigg(r^5 \varphi ^2 \chi
^2-2 r^3 \varphi ^2 \chi +4 \xi  \varphi  \sqrt{r^2 \chi }-4 \xi ^2
r \chi \bigg)}{4 \pi  \big(r^2 \chi +1\big)^2 \big(r^4 \varphi ^2
\chi -4 \xi ^2\big)}.
\end{eqnarray}
Substituting this value in Eq.\eqref{11}, we obtain
\begin{eqnarray}\nonumber
\varrho&=&\frac{1}{2} \bigg[\zeta -\frac{1}{\big(r^2 \chi +1\big)^2}
\bigg\{2 \chi  \psi  \big(r^2 \chi+3\big)\bigg\} +\bigg\{ \psi
\big(r^5 \varphi ^2 \chi ^2-2 r^3 \varphi ^2 \chi \\\label{15}&+&4
\xi \varphi \sqrt{r^2 \chi }-4 \xi ^2 r \chi \big)r
\chi\bigg\}\bigg\{\big(r^2 \chi +1\big)^2 \big(r^4 \varphi ^2 \chi
-4 \xi ^2\big)\bigg\}^{-1}\bigg].
\end{eqnarray}
Adding Eqs.\eqref{12} and \eqref{13}, it follows that
\begin{eqnarray}\nonumber p&=&-\bigg[4 \zeta  \xi  r^7 \varphi
\chi ^3-4 \xi  r^5 \varphi \chi ^2 (\chi  \psi -2 \zeta )+4 \xi  r^3
\varphi  \chi  (\zeta +\chi \psi )+\big(r^2 \chi
\big)^{\frac{3}{2}}\\\nonumber&\times& r^2\big(\zeta \big(4 \xi ^2
\chi +\varphi ^2\big)+4 \varphi ^2 \chi \psi \big)+4 \big(r^2 \chi
\big)^{\frac{3}{2}} \big(2 \zeta  \xi ^2-\xi ^2 \chi  \psi +2
\varphi ^2 \psi \big) \\\nonumber&+&4\xi ^2\sqrt{r^2 \chi } (\zeta
-2 \chi \psi )+r^2 \varphi ^2 \big(r^2 \chi \big)^{\frac{5}{2}} (2
\zeta -\chi \psi )+\zeta r^8 \varphi ^2 \chi ^3 \sqrt{r^2 \chi }
\\\label{14}&+&16 \xi r \varphi
\chi \psi \bigg]\bigg[2 \big(r^2 \chi +1\big)^2 \big(r^3 \varphi
\chi +2 \xi \sqrt{r^2 \chi }\big) \big(2 \xi +r \varphi \sqrt{r^2
\chi }\big)\bigg]^{-1}.
\end{eqnarray}

The scientific purpose of this study is to investigate charged
gravastar solutions within the novel framework of $f(\mathbb{Q})$
gravity using the Finch-Skea metric. It effectively explores the
impact of charge on the structural stability and physical properties
of gravastars, while also highlighting that the insights extend
beyond previous studies, which primarily focused on uncharged
models. By analyzing critical factors such as energy density,
entropy and proper length, we seek to demonstrate how charge
contributes to the overall stability of gravastars, enabling them to
avoid singularities and event horizons typically associated with
black holes.

\section{Structure of Charged Gravastar}

In this section, we analyze three distinct regions of a gravastar in
the presence of an electromagnetic field
\begin{itemize}
\item Interior Region: In this region, $p=-\varrho $,
\item Thin-Shell: The thin-shell represents $p=\varrho$,
\item Exterior Region: In the outer region of the gravastar, $p=0$.
\end{itemize}

\subsection{Interior Region}

The fundamental cosmic EoS parameter is expressed as
$\omega=\frac{p}{\varrho}$. This equation is followed by three
distinct zones in the foundational model \cite{1a}. In this context,
we assume the existence of an intriguing gravitational source within
the interior region. Although dark matter and DE are typically
viewed as separate phenomena, there is a possibility that they could
be different manifestations of the same entity. Our objective is to
investigate the EoS that characterizes the dark sector in the inner
region as
\begin{equation}\label{17}
p=-\varrho.
\end{equation}
By applying Eq.\eqref{17} to \eqref{14} and then adding the result
to \eqref{15}, we obtain
\begin{equation}\label{18}
\frac{\chi  \psi  \big(8 \xi  r^3 \varphi  \chi +4 \xi ^2 \sqrt{r^2
\chi }+3 r^2 \varphi ^2 \big(r^2 \chi \big)^{\frac{3}{2}}+2 r^2
\varphi ^2 \sqrt{r^2 \chi }+4 \xi  r \varphi \big)}{\big(r^2 \chi
+1\big) \big(r^3 \varphi  \chi +2 \xi  \sqrt{r^2 \chi }\big) \big(2
\xi +r \varphi  \sqrt{r^2 \chi }\big)}= 0.
\end{equation}
Using the above equation in Eq.\eqref{16}, we can derive an
expression for the charge as follows
\begin{eqnarray}\nonumber
\mathcal{E}^{2}&=& \frac{1}{\pi \big(4 \xi ^2-r^4 \varphi ^2 \chi
\big)^3}\bigg[2 r^2 \varphi ^2 \psi  \bigg(r^2 \varphi ^2 \bigg(8
\xi ^2 \chi  \big(r^2 \chi -2\big)\\\label{19}&+&\varphi ^2 \big(r^2
\chi \big(r^2 \chi \big(4-r^2 \chi \big)+16\big)+8\big)\bigg)-16 \xi
^4 \chi \bigg)\bigg].
\end{eqnarray}
Using Eq.\eqref{15} with the active gravitational mass
($\mathcal{M}(r)=4\pi\int_0^r\gamma^2\varrho(\gamma)d\gamma$), we
obtain
\begin{eqnarray}\nonumber
\mathcal{M}(r)&=& 4 \pi  \int_0^r \frac{1}{2} r^2 \bigg(\zeta
+\frac{1}{\big(r^2 \chi +1\big)^2 \big(r^4 \varphi ^2 \chi -4 \xi
^2\big)}\bigg[\chi \psi \bigg(4 \xi ^2 \big(r^2 \chi
+6\big)\\\label{20}&+&4 \xi r \varphi \sqrt{r^2 \chi }-r^4 \varphi
^2 \chi \big(r^2 \chi +8\big)\bigg)\bigg]\bigg) \, dr.
\end{eqnarray}

\subsection{Shell}

In this scenario, we use a stiff perfect fluid contained within the
thin-shell, which follows the EoS
\begin{equation}\label{19a}
p=\varrho.
\end{equation}
This particular EoS is a special case of a barotropic EoS with
$\omega=1$. Barotropic fluids are characterized by the relationship
$p=p(\varrho)$, meaning that the pressure depends only on the
density. Even though such fluids are considered unlikely in
practical situations, they are useful for theoretical studies
because they simplify the analysis and illustrate various approaches
to solve different problems. The concept of a stiff fluid was first
introduced by Zel$^{\prime}$dovich \cite{1b}, who associated it with
a cold baryonic universe. Staelens et al. \cite{1c} examined the
collapse of a spherical region of such a fluid within a cosmic
background. However, solving the field equations for this non-vacuum
region or shell proves to be quite challenging \cite{1d}. An
analytical solution was found under the conditions of thin-shell
approximation, where $0<e^{-\beta(r)}<<1 $. According to Israel
\cite{2a1}, the area separating the two spacetimes must be a
thin-shell. Additionally, any parameter that depends on $r$ can be
considered very small ($<< 1$) when $r$ approaches to 0. This
approximation allows us to simplify our equations, reducing them
from Eqs.\eqref{2}-\eqref{4} to the following form
\begin{eqnarray}\label{21}
\varrho+2 \pi \mathcal{E}^{2}&=&\frac{f}{2}- f_{\mathbb{Q}}\bigg[
\mathbb{Q}+\frac{e^{-\beta}}{r}\beta^{\prime}+\frac{1}{r^{2}}\bigg],
\\\label{22} p-2 \pi
\mathcal{E}^{2}&=&-\frac{f}{2}+ f_{\mathbb{Q}}\bigg(
\mathbb{Q}+\frac{1}{r^{2}}\bigg),
\\\label{23} p+2 \pi
\mathcal{E}^{2}&=& f_{\mathbb{Q}}
\bigg(\frac{\mathbb{Q}}{2}-e^{-\beta}
\bigg(\frac{-\alpha^{\prime}\beta^{\prime}}{4}-\frac{\beta^{\prime}}{2
r}\bigg)\bigg)-\frac{f}{2}.
\end{eqnarray}
Using Eqs.\eqref{5}, \eqref{6}, \eqref{10} and \eqref{19a} in
\eqref{21}-\eqref{23}, then adding Eqs.\eqref{21} and \eqref{22},
and also \eqref{22} and \eqref{23}. Solving the resulting two
equations, we obtain
\begin{equation}\label{24}
e^{-\beta (r)}=-c_{1}-\frac{\big(\zeta  r^2-2 \psi \big) \big(2 \xi
+r \varphi  \sqrt{r^2 \chi }\big)}{2 \psi  \big(2 \xi +5 r \varphi
\sqrt{r^2 \chi }\big)},
\end{equation}
where $c_{1}$ represents an integration constant.

\subsection{Exterior Region}

\subsubsection{Reissner-Nordstr$\ddot{o}$m Black Hole}

The outer region of the charged gravastar is believed to follow the
condition $p=0$, which means that the shell is entirely vacuum
sealed. This outer region is described by the
Reissner-Nordstr$\ddot{o}$m (RN) black hole as
\begin{equation}\label{25}
ds_+^{2}=-\bigg(1+\frac{Q^{2}}{r^{2}}-\frac{2M}{r}\bigg)dt^{2}
+\bigg(1+\frac{Q^{2}}{r^{2}}-\frac{2M}{r}\bigg)^{-1}dr^{2}
+r^{2}(d\theta^{2}+\sin^{2}\theta d\phi^{2}),
\end{equation}
where $``+"$ represents the exterior solution, while $M$ and $Q$
correspond to the mass and charge of the black hole, respectively.

\subsubsection{Regular Black Holes}

For the exterior geometry, we can select two types of regular black
holes to form the outer layer of the shell, aiming to address the
singularity problem by modifying their structures. The metric is
generally written as
\begin{equation}\label{26}
ds_+^{2}= -F(r)dt^{2}+F(r)^{-1}dr^{2}+
r^{2}(d\theta^{2}+\sin^{2}\theta d\phi^{2}).
\end{equation}
The Bardeen regular black hole is used for the outer layer because
it avoids singularities, meets the conditions of flatness as well as
weak energy condition and has a regular center. The regular Hayward
black hole is used to describe the exterior, as its metric was
developed to show how mass might accumulate around the Bardeen black
hole. For the Bardeen regular black hole, the function $F(r)$ is
given by
\begin{equation}\label{26a}
F(r)=1-\frac{2Mr^{2}}{(r^{2}+e^{2})^{\frac{3}{2}}}+\frac{Q^{2}}{r^{2}},
\end{equation}
while for the Hayward regular black hole, this is expressed as
\begin{equation}\label{26b}
F(r)=1-\frac{2Mr^{2}}{r^{3}+2Ml^{2}}+\frac{Q^{2}}{r^{2}}.
\end{equation}
In this context, $e$ is the magnetic charge parameter and $l$ is a
constant that relates to the size of the core region. We note that
the Bardeen black hole reduces to the Schwarzschild black hole when
both $e=0$ and $Q=0$, while it reduces to the RN black hole for
$e=0$. Similarly, the Hayward black hole reduces to the
Schwarzschild black hole when $l =0$ and $Q =0$, while it reduces to
the RN black hole for $l =0$.

\section{Boundary and Junction Conditions}

In this section, we align our internal spacetime with the external
to determine the constants $\chi$, $\xi$ and $\varphi$. To do this,
we establish a set of relationships at the boundary surface where
$r=R$, which separates the inner and outer regions. These
relationships are derived from the requirement that the metric
coefficients $g_{tt}$ and $g_{rr}$ remain continuous across the
boundary. We also ensure that the derivative of $g_{tt}$ with
respect to $r$ is continuous. This matching process is crucial for
maintaining the smoothness and consistency of the spacetime geometry
at the interface between the two regions. In the following, we
calculate the values of constants $\chi$, $\xi$ and $\varphi$ for
different black holes.

First, we calculate these constants for the RN black hole. Here, we
compare the temporal and radial metric coefficients from
Eqs.\eqref{1} and \eqref{10} and use \eqref{25}, it follows that
\begin{eqnarray}\label{27}
\frac{Q^2}{R^2}-\frac{2 M}{R}+1&=&\bigg(\xi +\frac{1}{2} R \varphi
\sqrt{R^2 \chi }\bigg)^2,\\\label{28} \bigg(\frac{Q^2}{R^2}-\frac{2
M}{R}+1\bigg)^{-1}&=&R^2 \chi +1.
\end{eqnarray}
Differentiating Eq.\eqref{27} with respect to $R$, we obtain
\begin{eqnarray}\label{29}
\frac{M}{R^2}-\frac{Q^2}{R^3}&=&\frac{1}{2} \varphi \bigg(R^3
\varphi \chi +2 \xi  \sqrt{R^2 \chi }\bigg).
\end{eqnarray}
The constants $\chi$, $\xi$, and $\varphi$ are determined by solving
Eqs.\eqref{27}-\eqref{29} simultaneously as
\begin{eqnarray}\label{30}
\chi &=&\frac{-Q^2+2 R M }{R^2(Q^2+R^2-2 R M)}, \\\label{31}
\xi&=&\frac{3 Q^2-5 R M+2 R^2}{2 R \sqrt{Q^2+R^2-2 R
M}},\\\label{32} \varphi&=&\frac{-Q^2+R M}{R^2 \sqrt{-Q^2+2 R M}}.
\end{eqnarray}
Using the previous procedure, we can find the constants for the
Bardeen black hole
\begin{eqnarray}\label{33}
\frac{Q^2}{R^2}-\frac{2 M
R^2}{\big(e^2+R^2\big)^{\frac{3}{2}}}+1&=&\bigg(\xi +\frac{1}{2} R
\varphi \sqrt{R^2 \chi }\bigg)^2, \\\label{34}
\bigg(\frac{Q^2}{R^2}-\frac{2 M
R^2}{\big(e^2+R^2\big)^{\frac{3}{2}}}+1\bigg)^{-1}&=&R^2 \chi +1.
\end{eqnarray}
\begin{eqnarray}\label{35}
\frac{3 M R^3}{(e^2 + R^2)^\frac{5}{2}}-\frac{Q^2}{R^3}- \frac{2 M
R}{(e^2 + R^2)^\frac{3}{2}}&=&\frac{1}{2} \varphi \bigg(R^3 \varphi
\chi +2 \xi \sqrt{R^2 \chi }\bigg).
\end{eqnarray}
Solving this system of equations, we obtain
\begin{eqnarray}\label{36}
\chi&=&\frac{\frac{2MR^2}{\big(e^2+R^2\big)^{\frac{3}{2}}}-\frac{Q^2}{R^2}}{R^2
\bigg(\frac{Q^2}{R^2}-\frac{2MR^2}{\big(e^2+R^2\big)^{\frac{3}{2}}}+1\bigg)},
\\\nonumber
\xi &=&\bigg[\bigg[\bigg\{R^3 \varphi  \big(e^2+R^2\big) \big(e^4
Q^2+R^4 \big(Q^2-2 M \sqrt{e^2+R^2}\big)+2 e^2 Q^2
R^2\big)\bigg\}\\\nonumber&\times&\bigg\{\bigg(-\bigg\{e^4 Q^2+R^4
\big(Q^2-2 M \sqrt{e^2+R^2}\big)+2 e^2 Q^2
R^2\bigg\}\bigg\{\big(Q^2+R^2\big)\\\nonumber&\times&2 e^2 R^2 +R^4
\big(Q^2-2 M \sqrt{e^2+R^2}+R^2\big)+e^4
\big(Q^2+R^2\big)\bigg\}^{-1}\bigg)^{\frac{1}{2}}\bigg\}^{-1}\bigg]\\\nonumber&+&2
\bigg(R^2 \big(e^2+R^2\big) \bigg(e^{10} \big(Q^2+R^2\big)^3+5 e^8
R^2 \big(Q^2+R^2\big)^3+\big(Q^2+R^2\big)\\\nonumber&\times& e^2 R^8
\big(\big(Q^2+R^2\big) \big(5 Q^2-18 M \sqrt{e^2+R^2}+5 R^2\big)+24
M^2 R^2\big)+R^{10}\\\nonumber&\times& \bigg(3 Q^4 \big(R^2-2 M
\sqrt{e^2+R^2}\big)-8 M^3 R^2 \sqrt{e^2+R^2}+12 M^2 R^4+Q^6
\\\nonumber&+&\big(4 M \big(M-\sqrt{e^2+R^2}\big)+R^2\big)3 Q^2 R^2-6 M R^4
\sqrt{e^2+R^2}+R^6\bigg)\\\nonumber&+&2 e^6 R^4 \big(Q^2+R^2\big)^2
\big(5 R^2-3 M \sqrt{e^2+R^2}+5 Q^2\big)+2 e^4
\big(Q^2+R^2\big)\\\nonumber&\times& R^6\bigg(\big(Q^2+R^2\big)
\big(5 Q^2-9 M \sqrt{e^2+R^2}+5 R^2\big)+6 M^2
R^2\bigg)\bigg)\bigg)^{\frac{1}{2}}\bigg]\\\nonumber&\times&\bigg[2
R^2 \big(e^2+R^2\big) \bigg(e^4 \big(Q^2+R^2\big)+R^4 \big(-2 M
\sqrt{e^2+R^2}+Q^2+R^2\big)\\\label{37}&+&2 e^2 R^2
\big(Q^2+R^2\big)\bigg)\bigg]^{-1},\\\nonumber \varphi
&=&-\bigg[\big(e^2+R^2\big) \bigg(e^4 \big(Q^2+R^2\big)+R^4
\big(Q^2-2 M \sqrt{e^2+R^2}+R^2\big)+2 e^2\\\nonumber&\times& R^2
\big(Q^2+R^2\big)\bigg) \bigg(4 l^4 M^2 Q^2+4 l^2 M R^3 \big(M
R+Q^2\big)+R^6\big(Q^2-MR\big)\bigg)\bigg]\\\nonumber&\times&\bigg[R
\big(2 l^2 M+R^3\big)^2 \bigg\{\bigg(\frac{Q^2}{R^2}-\frac{2 M
R^2}{\big(e^2+R^2\big)^{3/2}}+1\bigg)^{-1}-1\bigg\}^{\frac{1}{2}}
\bigg\{R^2 \big(e^2+R^2\big)\\\nonumber&\times&\bigg(e^{10}
\big(Q^2+R^2\big)^3+5 e^8 R^2 \big(Q^2+R^2\big)^3+e^2 R^8
\big(Q^2+R^2\big) \bigg(\big(Q^2+R^2\big)\\\nonumber&\times&\big(5
Q^2-18 M \sqrt{e^2+R^2}+5 R^2\big)+24 M^2 R^2\bigg)+R^{10} \bigg(3
Q^4- \sqrt{e^2+R^2}\\\nonumber&\times&8 M^3 R^2 \big(R^2-2 M
\sqrt{e^2+R^2}\big)+3 Q^2 R^2 \big(4 M
\big(M-\sqrt{e^2+R^2}\big)+R^2\big)\\\nonumber&-&6 M R^4
\sqrt{e^2+R^2}+12 M^2 R^4+Q^6+R^6\bigg)+2 e^6 R^4
\big(Q^2+R^2\big)^2 \big(5 Q^2\\\nonumber&-&3 M \sqrt{e^2+R^2}+5
R^2\big)+2 e^4 R^6 \big(Q^2+R^2\big) \bigg( \big(5 Q^2-9 M
\sqrt{e^2+R^2}\\\label{38}&+&5 R^2\big)\big(Q^2+R^2\big)+6 M^2
R^2\bigg)\bigg)\bigg\}^{\frac{1}{2}}\bigg]^{-1}.
\end{eqnarray}
Similarly, we can obtain the values of constants for Hayward black
hole
\begin{align}\label{42}
\chi &=\frac{\frac{2 M R^2}{2 l^2 M+R^3}-\frac{Q^2}{R^2}}{R^2
\left(-\frac{2 M R^2}{2 l^2 M+R^3}+\frac{Q^2}{R^2}+1\right)},
\\\nonumber
\xi&=\bigg[2 R^2 \big(2 l^2 M+R^3\big) \big(2 l^2 M
\big(Q^2+R^2\big)+R^3 \big(R (R-2
M)+Q^2\big)\big)\bigg]^{-1}\bigg[\bigg\{R^2
\\\nonumber
&\times \big(2 l^2 M+R^3\big) \big(2 l^2 M \big(Q^2+R^2\big)+R^3
\big(R (R-2 M)+Q^2\big)\big)^3\bigg\}^{\frac{1}{2}}2-\bigg(R^3
\\\nonumber
&\times \varphi \big(2 l^2 M+R^3\big) \big(R^3 \big(2 M R-Q^2\big)
-2 l^2 M Q^2\big)\bigg)\bigg(\bigg\{\bigg(+ R^3 \big(2 M R-Q^2\big)
\\\label{43}
&-2 l^2 M Q^2\bigg)\bigg(2 l^2 M \big(Q^2+R^2\big)+R^3 \big(R (R-2
M)+Q^2\big)\bigg)^{-1}\bigg\}^{\frac{1}{2}}\bigg)^{-1}\bigg],
\\\nonumber
\varphi &=-\bigg[\big(2 l^2 M R+R^4\big)\bigg\{R^2 \big(2 l^2
M+R^3\big)\big(2 l^2 M \big(Q^2+R^2\big) +R^3 \big(R (R-2 M)
\\\nonumber
&+Q^2\big)\big)^3\bigg\}^{\frac{1}{2}}\bigg(\bigg(\frac{Q^2}{R^2}-\frac{2
M R^2}{2 l^2 M+R^3}+1\bigg)^{-1}-1\bigg)^{\frac{1}{2}}\bigg]^{-1}
\bigg[\big(4 l^4 M^2 Q^2+4 l^2 M
\\\nonumber
&\times R^3 \big(M R+Q^2\big)+R^6 \big(Q^2-M R\big)\big) \big(2 l^2
M \big(Q^2+R^2\big)+R^3 \big(R (R-2 M)
\\\label{44}
&+Q^2\big)\big)\bigg].
\end{align}

Sen \cite{2b1} was the first who explored the junction conditions.
Junction conditions are rules that describe how two different
regions, like the inside and outside of a star, connect smoothly.
Darmois junction conditions \cite{2d1} ensure that this connection
is smooth by requiring the following: the surface where the regions
meet must have the same shape and size on both sides (continuity of
the metric) and the way the surface curves should be consistent from
both sides (continuity of the extrinsic curvature). Israel junction
conditions \cite{2a1} apply when there is a thin layer at the
boundary such as a membrane, which can have its own energy and
pressure. In this case, Israel conditions allow for a sudden change
in how the surface curves across this layer, which is related to the
properties of the thin layer itself.

Here, we adopt Israel junction conditions \cite{3a1} due to their
suitability for scenarios involving a boundary between two regions
of spacetime with a thin-shell of matter or energy. This approach is
particularly useful when dealing with surface layers of mass and
charge. While the metric coefficients remain continuous across the
boundary, their derivatives may exhibit discontinuities at the
hypersurface, which aligns well with the conditions necessary for
analyzing such thin-shells.

The line element of the induced metric at the boundary is given by
the following expression
\begin{equation}
ds^2 = -d\tau^2 + R(\tau)^2 d\theta^2 + R(\tau)^2 \sin^2\theta \,
d\phi^2,
\end{equation}
where $\tau$ is the proper time. To ensure that the thin-shell
remains stable, we need to use the Lanczos equation to calculate the
surface tension and pressure
\begin{equation}\label{50}
S^{i}_{j}= -\frac{1}{8\pi}\{[K^{i}_{j}]-\delta^{i}_{j}K\},
\end{equation}
here, $i$ and $j$ are coordinates on the hypersurface $(i,j=0,2,3)$
and $[K^{i}_{j}] = K^{+i}_{j}-K^{-i}_{j}$,
$K=$~tr$[K_{ij}]=[K^{i}_{j}]$ represents the discontinuity in the
extrinsic curvature. The extrinsic curvature is defined as
\begin{equation}\label{48}
K_{ij}^{\pm}=-n_{v}^{\pm}\bigg(\frac{d^{2}x^{v}_{\pm}}
{d\phi^{i}d\phi^{j}}+ \Gamma_{lm}^{v}\frac{d
x^{l}_{\pm}}{d\phi^{i}}\times\frac{dx^{m}_{\pm}}{d\phi^{j}}\bigg),\quad
l,m=0,1,2,3,
\end{equation}
where $\phi^{i}$ are the intrinsic coordinates and hence, the
components of $K^{i\pm}_{j}$ are given by
\begin{equation}\label{4aab}
K^{\tau\pm}_{\tau} = \frac{\Phi^\prime +
2\ddot{R}}{\sqrt{\Phi_{\pm}(R) + \dot{R}^2}}, \quad
K^{\theta\pm}_{\theta} = \frac{\sqrt{\Phi_{\pm}(R) + \dot{R}^2}}{R},
\quad K^{\phi\pm}_{\phi} = \sin^2\theta \, K^{\theta\pm}_{\theta}.
\end{equation}
The unit normals to the hypersurface ($n_{v}^{\pm}$) are
\begin{equation}\label{49}
n_{v}^{\pm}=\bigg(\frac{\dot{R}}{\Phi_{\pm}(R)},
\sqrt{\Phi_{\pm}(R)+\dot{R}^{2}},0,0\bigg),
\end{equation}
where $\dot{R}$ represents derivative of $R$ with respect to $\tau$.
The surface energy density $\rho$ and surface pressure $P$ of
thin-shell gravastars can be determined using the Lanczos equations
as
\begin{eqnarray}\label{4aac}
\rho&=& -\frac{[K^\theta_\theta]}{4\pi} = -\frac{1}{4\pi
R}\left(\sqrt{\dot{R}^2 + \Phi_{+}(R)} - \sqrt{\dot{R}^2 +
\Phi_{-}(R)}\right),
\\\nonumber P&=& \frac{[K^\theta_\theta] + [K^\tau_\tau]}{8\pi}=
\frac{1}{8\pi R}\bigg[\frac{2\dot{R}^2 + 2R\ddot{R} + 2\Phi_{+}(R) +
R\Phi^{\prime}_{+}(R)}{\sqrt{\dot{R}^2 + \Phi_{+}(R)}}
\\\label{4aad}&-&\frac{2\dot{R}^2 + 2R\ddot{R}
+ 2\Phi_{-}(R) + R\Phi^{\prime}_{-}(R)}{\sqrt{\dot{R}^2 +
\Phi_{-}(R)}}\bigg].
\end{eqnarray}

At equilibrium, when the shell is stationary at the throat radius $R
= R_{0}$ (i.e., $\dot{R_{0}} = 0$ and $\ddot{R_{0}}=0$), these
equations simplify to
\begin{eqnarray}\label{51}
\rho_{0}&=&-\frac{1}{4\pi
R_{0}}[\sqrt{\Phi_{+}(R_{0})}-\sqrt{\Phi_{-}(R_{0})}],
\\\label{52}
P_{0}&=&-\rho_{0}
+\frac{1}{8\pi}\bigg[\frac{\Phi_{+}^{\prime}(R_{0})}{\sqrt{\Phi_{+}(R_{0})}}-
\frac{\Phi_{-}^{\prime}(R_{0})}{\sqrt{\Phi_{-}(R_{0})}}\bigg].
\end{eqnarray}
The exterior function of RN black hole is defined as
\begin{equation}\label{52a}
\Phi_{+}(R_{0})=1-\frac{2 M}{R_{0}}+\frac{Q^2}{R_{0}^2},
\end{equation}
and the interior function is described as
\begin{eqnarray}\label{52b}
\Phi_{-}(R_{0})&=&\frac{\big(2 R_{0}(R_{0}-2 M)+Q^2\big) \big(\zeta
R_{0}^2-2 \psi \big)}{6 Q^2 \psi -4 R_{0}^2 \psi }-c_{1}.
\end{eqnarray}
Similarly, the exterior function of Bardeen black hole is defined as
\begin{equation}\label{52c}
\Phi_{+}(R_{0})=\frac{Q^2}{R_{0}^2}-\frac{2 M
R_{0}^2}{\big(e^2+R_{0}^2\big)^{\frac{3}{2}}}+1,
\end{equation}
and the interior function is
\begin{eqnarray}\nonumber
\Phi_{-}(R_{0})&=&\big(-\big((R_{0}^2 \zeta -2 \psi)(\big((2 l^2
M+R_{0}^3)^2 \big(R_{0}(e^2+R_{0}^2)(e^4(Q^2+R_{0}^2)+R_{0}^4
\\\nonumber &\times&(-2 M(e^2+R_{0}^2)^{\frac{1}{2}}+Q^2+R_{0}^2)+2 e^2
R_{0}^2(Q^2+R_{0}^2))^2(-4 l^4 M^2 Q^2\\\nonumber &-&4 l^2M R_{0}^3
(M R_{0}+Q^2)+R_{0}^6(M R_{0}-Q^2)) ((R_{0}^2(2 M
R_{0}^4-(e^2+R_{0}^2)^{\frac{3}{2}}\\\nonumber &\times&Q^2))(R_{0}^2
(e^2+R_{0}^2)^{\frac{3}{2}}(Q^2+R_{0}^2)-2 M
R_{0}^6)^{-1})^{\frac{1}{2}}\big)\big(\big(-(2 M R_{0}^2)((e^2
\\\nonumber
&+&R_{0}^2)^{3/2})^{-1}+\frac{Q^2}{R_{0}^2}+1\big)^{-1}
-1\big)^{\frac{1}{2}}\big)^{-1}-\big(\big(R_{0}(e^2+R_{0}^2)(2 l^2
M+R_{0}^3)^2\\\nonumber &\times&(e^4(Q^2 +R_{0}^2)+R_{0}^4(Q^2-2 M
\sqrt{e^2+R_{0}^2}+R_{0}^2)+2 e^2 R_{0}^2(Q^2+R_{0}^2))^2\\\nonumber
&\times&(R_{0}^6(M R_{0}-Q^2)-4 l^4 M^2 Q^2-4
l^2(MR_{0}+Q^2))\big)((2 l^2 M
+R_{0}^3)^2)^{-1}\big)^{-1}\\\nonumber
&+&\big((e^{10}(Q^2+R_{0}^2)^3+5 e^8 R_{0}^2(Q^2+R_{0}^2)^3+e^2
R_{0}^8 (Q^2+R_{0}^2) ((Q^2+R_{0}^2)
\\\nonumber
&\times&(5 Q^2-18 M \sqrt{e^2+R_{0}^2}+5 R_{0}^2)+24 M^2
R_{0}^2)+R_{0}^{10}(-8 M^3 R_{0}^2 \sqrt{e^2+R_{0}^2}
\\\nonumber
&+&3 Q^4(R_{0}^2-2 M\sqrt{e^2+R_{0}^2})+3 Q^2 R_{0}^2(4 M
(M-\sqrt{e^2+R_{0}^2})+R_{0}^2)-6
\\\nonumber
&\times& M R_{0}^4\sqrt{e^2+R_{0}^2}+12 M^2 R_{0}^4+Q^6+R_{0}^6)+2
e^6 R_{0}^4 (Q^2+R_{0}^2)^2( R_{0}^2-3
\\\nonumber
&\times&M \sqrt{e^2+R_{0}^2}+5 Q^2)+2 e^4 R_{0}^6
(Q^2+R_{0}^2)((Q^2+R_{0}^2)(Q^2- M \sqrt{e^2+R_{0}^2}
\\\nonumber
&+&R_{0}^2)+6 M^2 R_{0}^2))\big)\big)\big)\big(2 \psi(5 \big(R_{0}
(e^2+R_{0}^2)(e^4(Q^2+R_{0}^2)+R_{0}^4(Q^2+R_{0}^2)
\\\nonumber
&+& e^2 R_{0}^2(Q^2+R_{0}^2))^2(4 l^4 M^2 Q^2-4 l^2 M R_{0}^3(M
R_{0}+Q^2)+R_{0}^6(M R_{0}-Q^2))\\\nonumber &\times& ((R_{0}^2(
R_{0}^4-Q^2(e^2+R_{0}^2)^{3/2}))(R_{0}^2
(e^2+R_{0}^2)^{3/2}(Q^2+R_{0}^2)- M
R_{0}^6)^{-1})^{\frac{1}{2}}\big)\\\nonumber &\times&((2 l^2
M+R_{0}^3)^2 (\big(-\frac{2 M
R_{0}^2}{(e^2+R_{0}^2)^{3/2}}+\frac{Q^2}{R_{0}^2}+1\big)^{-1}-1)^{\frac{1}{2}})^{-1}
-\big((e^2+R_{0}^2)\\\nonumber &\times&R_{0}(2 l^2
M+R_{0}^3)^2(e^4(Q^2 +R_{0}^2)+R_{0}^4(-2 M
\sqrt{e^2+R_{0}^2}+Q^2+R_{0}^2)+2
\\\nonumber
&\times&e^2 R_{0}^2(Q^2+R_{0}^2))^2(4 l^4 M^2 Q^2-4 l^2 M R_{0}^3(M
R_{0}+Q^2)+R_{0}^6(M R_{0}-Q^2))\big)
\\\nonumber
&\times&(2 l^2 M+R_{0}^3)^2)^{-1}+\big(2
R_{0}(e^{10}(Q^2+R_{0}^2)^3+5 e^8 R_{0}^2(Q^2+R_{0}^2)^3+e^2 R_{0}^8
\\\nonumber
&\times&(-18 M \sqrt{e^2+R_{0}^2}+5 Q^2+5 R_{0}^2)+24 M^2
R_{0}^2)+R_{0}^{10}(-8 M^3 R_{0}^2 \sqrt{e^2+R_{0}^2}
\\\nonumber
&+&3 Q^4(R_{0}^2-2 M \sqrt{e^2+R_{0}^2})+3 Q^2 R_{0}^2(4 M
(M-\sqrt{e^2+R_{0}^2})+R_{0}^2)-6 M R_{0}^4
\\\nonumber
&\times& \sqrt{e^2+R_{0}^2}+12 M^2 R_{0}^4+Q^6+R_{0}^6)+2 e^6
R_{0}^4 (Q^2+R_{0}^2)^2(-3 M \sqrt{e^2+R_{0}^2}
\\\nonumber
&+&5 Q^2+5 R_{0}^2)+2 e^4 R_{0}^6 (Q^2+R_{0}^2)((Q^2+R_{0}^2)(-9 M
\sqrt{e^2+R_{0}^2}+5 Q^2
\\\label{52d}
&+&5 R_{0}^2)+6 M^2
R_{0}^2))\big)\big)\big)^{-1}-c_{1}\big)^{\frac{1}{2}}\bigg].
\end{eqnarray}
Finally, the exterior function of Hayward black hole is given as
\begin{equation}\label{52e}
\Phi_{+}(R_{0})=\frac{Q^2}{R_{0}^2}-\frac{2 M R_{0}^2}{2 l^2
M+R_{0}^3}+1,
\end{equation}
while the interior function is
\begin{eqnarray}\nonumber
\Phi_{-}(R_{0})&=&\bigg\{-(R_{0}(-4 l^4 M^2 Q^2-4 l^2 M R_{0}^3(M
R_{0}+Q^2)+R_{0}^6(M R_{0}-Q^2))
\\\nonumber
&\times&(2 l^2 M (Q^2+R_{0}^2)+R_{0}^3(R_{0} (R_{0}-2
M)+Q^2))^2\bigg\{R_{0}^2
\bigg(\bigg(\frac{Q^2}{R_{0}^2}+1\\\nonumber &-&\frac{2 M R_{0}^2}{2
l^2
M+R_{0}^3}\bigg)^{-1}-1\bigg)\bigg\}(R_{0}^2)^{-1})^{\frac{1}{2}}(R_{0}^2
\zeta -2 \psi ))((2 l^2 M R_{0}+R_{0}^4)
\\\nonumber
&\times&(R_{0}^2(2 l^2 M+R_{0}^3)(2 l^2 M (Q^2+R_{0}^2)+R_{0}^3
(R_{0} (R_{0}-2 M)+Q^2))^3)^{\frac{1}{2}}
\\\nonumber
&\times&(\frac{Q^2}{R_{0}^2}+1-(2 M R_{0}^2)(2 l^2
M+R_{0}^3))^{-1}-1)^{\frac{1}{2}})^{-1}+(R_{0}^2(2 l^2 M
\\\nonumber
&\times&(Q^2+R_{0}^2)+R_{0}^3(R_{0} (R_{0}-2 M)+Q^2))^5(4 l^4 M^2 (3
Q^2+2 R_{0}^2)\\\nonumber &+& 4 l^2 M R_{0}^3(R_{0} (2 R_{0}-M)+3
Q^2)+R_{0}^6(R_{0} (2 R_{0}-5 M)+3 Q^2)))((R_{0}^2\\\nonumber
&\times&(2 l^2 M+R_{0}^3)(2 l^2 M(Q^2+R_{0}^2)+R_{0}^3(R_{0}
(R_{0}-2 M) +Q^2))^3)^{3/2})^{-1}
\\\nonumber
&\times&(2 \bigg((R_{0}^2((Q^2+R_{0}^2)+4 l^2 M R_{0}^3(R_{0} (2
R_{0}-M)+3 Q^2)+R_{0}^6(R_{0} (2 R_{0}\\\nonumber &-&5 M)+3
Q^2)))((R_{0}^2(2 l^2 M +R_{0}^3)(2 l^2 M(Q^2+R_{0}^2)+R_{0}^3(R_{0}
(R_{0}-2
\\\nonumber &\times& M)+Q^2))^3)^{3/2})^{-1}+((5 (R_{0}(-4 l^4M^2 Q^2-4
l^2 M R_{0}^3(M R_{0}+Q^2)\\\nonumber &+&R_{0}^6(M R_{0}-Q^2))(2 l^2
M (Q^2+R_{0}^2) +R_{0}^3(R_{0} (R_{0}-2 M)+Q^2))^2\\\nonumber
&\times&\bigg\{R_{0}^2 \bigg((-\frac{2 M R_{0}^2}{2 l^2 M+R_{0}^3}
+\frac{Q^2}{R_{0}^2}+1)^{-1}-1\bigg)\bigg\}((2 l^2 M
R_{0}+R_{0}^4)(R_{0}^2
\\\nonumber
&\times&(2 l^2 M+R_{0}^3)(2 l^2 M (Q^2+R_{0}^2)+R_{0}^3 (R_{0}
(R_{0}-2 M)+Q^2))^3)^{\frac{1}{2}}
\\\label{52f}
&+&(-(2 M R_{0}^2)(2 l^2
M+R_{0}^3)+\frac{Q^2}{R_{0}^2}+1)^{-1}-1)^{\frac{1}{2}}\bigg)
-c_{1})^{-1}\bigg\}^{\frac{1}{2}}.
\end{eqnarray}

The dynamic properties of a thin-shell can be identified using the
equation of motion and the conservation equation. These equations
are crucial for understanding the stable regions of a geometrical
structure. First, we derive the equation of motion by reorganizing
Eq.\eqref{4aac} as
\begin{equation}\label{5abc}
\dot{R}^2 + V(R) = 0,
\end{equation}
where $V(R)$ represents the effective potential, which can be
written as
\begin{equation}\label{61}
V(R)=\frac{1}{2} \big( \Phi_{-}(R) + \Phi_{+}(R) \big) -
\frac{\big(\Phi_{-}(R) - \Phi_{+}(R) \big)^2}{64\pi^2 R^2 \rho^2} -
4\pi^2 R^2 \rho^2.
\end{equation}
Secondly, we analyze that $\rho$ and $P$ adhere to the conservation
equation
\begin{equation}\label{59}
\frac{d}{d\tau}(\rho 4\pi R^2)+P\frac{d 4\pi R^2}{d\tau}=0,
\end{equation}
Using the above equation, we can find
\begin{equation}\label{60}
\rho^{\prime}=-(\rho+P)\frac{2}{R}.
\end{equation}

Now we check the behavior of $\rho,~P$ at the junction surface
$R=R_{0}$ for three different black holes. For RN black hole, using
Eqs.\eqref{52a} and \eqref{52b} into \eqref{51} and \eqref{52}, we
have
\begin{eqnarray}\label{53}
\rho_{0}&=&\frac{\sqrt{-\frac{2
M}{R_{0}}+\frac{Q^2}{R_{0}^2}+1}-\sqrt{\frac{\big(2 R_{0} (R_{0}-2
M)+Q^2\big) \big(\zeta  R_{0}^2-2 \psi \big)}{6 Q^2 \psi -4 R_{0}^2
\psi }-c_{1}}}{4 \pi R_{0}},
\\\nonumber
P_{0}&=& \sqrt{1-\frac{2
M}{R_{0}}+\frac{Q^2}{R_{0}^2}}-\sqrt{\frac{\big(2 R_{0} (R_{0}-2
M)+Q^2\big) \big(\zeta  R_{0}^2-2 \psi \big)}{6 Q^2 \psi -4 R_{0}^2
\psi }-c_{1}}\\\nonumber&+&\frac{1}{8 \pi}\bigg(\frac{1}{R_{0}^2 }(M
R_{0}-Q^2\sqrt{R_{0}^2-2 M R_{0}+Q^2})+\bigg(2 Q^2 \big(9 \zeta M
R_{0}^2-6 M
\\\nonumber&\times&\psi -6\zeta R_{0}^3+8 R_{0} \psi \big)-4 M \big(\zeta
R_{0}^4+2 R_{0}^2 \psi \big)-3 \zeta Q^4 R_{0}+4 \zeta
R_{0}^5\bigg)\bigg(\sqrt{2}
\\\nonumber&\times&\psi \big(3 Q^2-2
R_{0}^2\big)^2 \bigg(\bigg(\big(2 c R_{0} \psi-2 \zeta M R_{0}^2+4 M
\psi +\zeta R_{0}^3-2 R_{0} \psi \big)2 R_{0} \\\label{54}&+&Q^2
\big(\zeta R_{0}^2-2 (3 c_{1} \psi +\psi )\big)\bigg)\bigg(\psi
\big(3 Q^2-2
R_{0}^2\big)\bigg)^{-1}\bigg)^{\frac{1}{2}}\bigg)^{-1}\bigg).
\end{eqnarray}
Similarly, we can find $\rho_{0}$ and $P_{0}$ for Bardeen and
Hayward black holes. However, due to complicated expressions, we are
only giving the graphical representation.

Figures \textbf{1-3} show similar behavior for the three black
holes. These indicate that the shell's outer boundary is denser than
its inner boundary due to the increasing energy density and pressure
for three different values of $Q$, $e$ and $l$. It is mentioned here
that we have taken $\zeta=2,~\psi=1$ in all graphs as these are the
only suitable values for the required physical behavior. These
figures highlight the transition of physical properties from the
inner to the outer boundary, demonstrating that energy density and
pressure increase as we approach the outer boundary. This results in
a denser outer boundary compared to the inner one. The variations
are influenced by the repulsive force introduced by charge $Q$ and
the electric field, which becomes more pronounced with increasing
charge, particularly near the outer boundary.
\begin{figure}\center
\epsfig{file=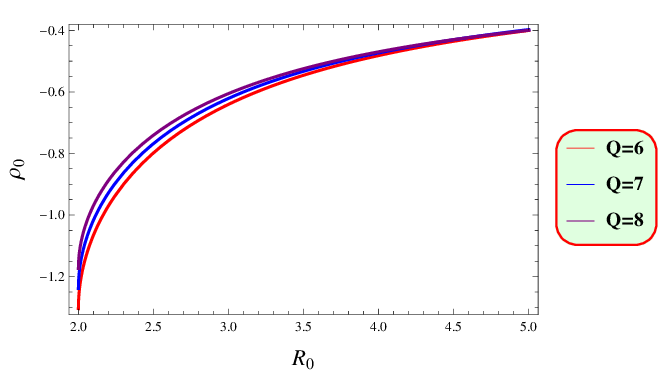,width=.47\linewidth}
\epsfig{file=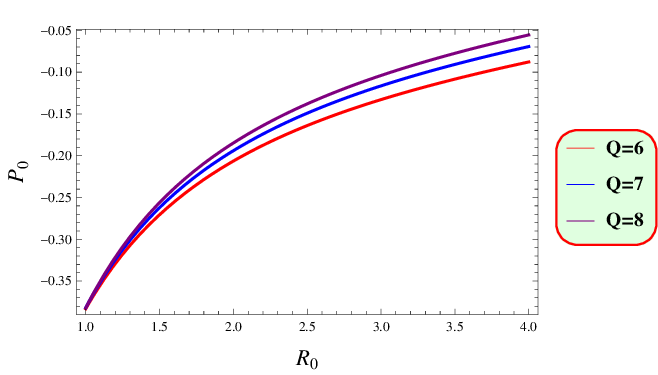,width=.47\linewidth} \caption{Plots of
$\rho_{0}$ and $P_{0}$ for the RN Black Hole versus $R_{0}$.}
\end{figure}
\begin{figure}\center
\epsfig{file=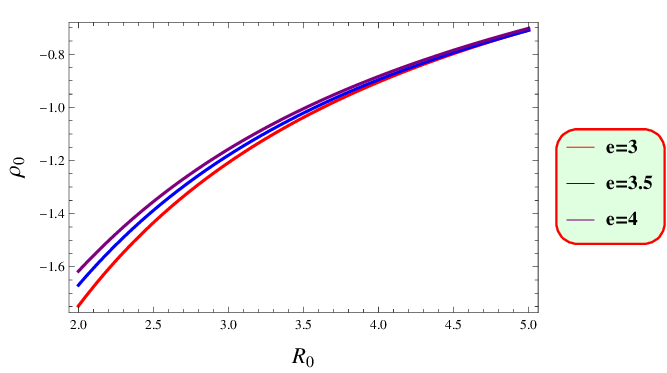,width=.47\linewidth}
\epsfig{file=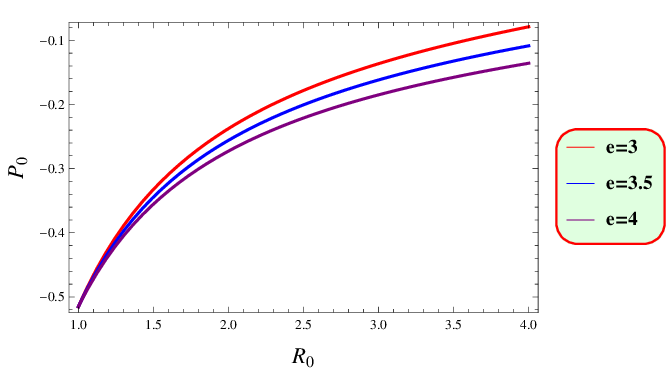,width=.47\linewidth} \caption{Graphs of
$\rho_{0}$ and $P_{0}$ for the Bardeen Black Hole versus $R_{0}$.}
\end{figure}
\begin{figure}\center
\epsfig{file=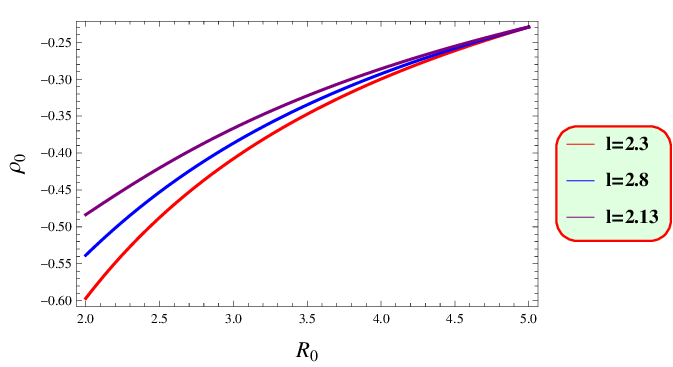,width=.47\linewidth}
\epsfig{file=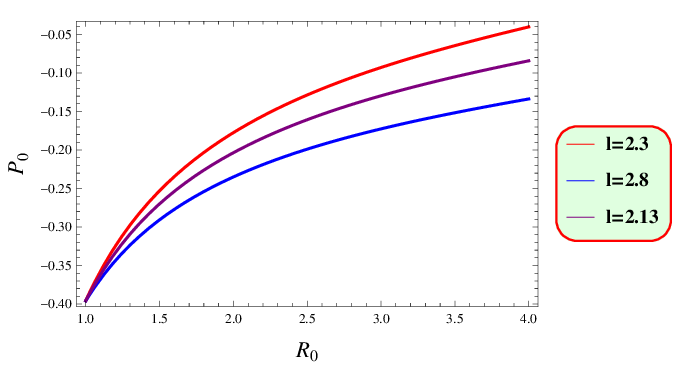,width=.47\linewidth} \caption{Graphs of
$\rho_{0}$ and $P_{0}$ for the Hayward black hole versus $R_{0}$.}
\end{figure}

\section{Physical Features of Charged Gravastar}

This section will cover the physical properties of charged gravastar
such as surface energy density, surface pressure, proper length,
entropy, energy, the EoS parameter. We also discuss stability of the
gravastar through the effective potential, causality condition and
adiabatic index. We shall restrict ourselves to RN black hole only
as the rest of the black holes provide complicated expressions and
cannot be analyzed.

\subsection{Proper length}

According to Mazur and Mottola \cite{1a}, the shell is situated at
the boundary where two different spacetimes intersect. The shell
stretches from the phase boundary at $r_2 = R_{0} + \epsilon$, which
separates the outer space-time from the intermediate thin-shell, to
the phase barrier at $r_1 = R_{0}$, which is located between the
inner region and the intermediate thin-shell ( $\epsilon$ is the
thickness of the shell and a small distance). The thickness of the
shell is determined between these boundary points as \cite{8aa}
\begin{equation}\nonumber
L = \int_{R_{0}}^{R_{0}+\varepsilon} \sqrt{e^{\beta(R)}} dR.
\end{equation}
This expression can be converted into the interior function as
\begin{eqnarray}\label{65a} L&=&\int_{R_{0}}^{R_{0}+\varepsilon}
\bigg(\frac{\big(2 R(R-2 M)+Q^2\big) \big(\zeta R^2-2 \psi \big)}{6
Q^2 \psi -4 R^2 \psi }-c_{1}\bigg)^{-1} dr,
\end{eqnarray}which
can also be written as (using Eq.\eqref{52b})
\begin{equation}\nonumber
L= \int_{R_{0}}^{R_{0}+\varepsilon} \frac{1}{\Phi_-(R)} dR.
\end{equation}
Evaluating the integral in Eq.\eqref{65a} is a bit challenging. To
simplify the process, we take a particular assumption that $\frac{d
\Phi_-(R)}{dR}=\frac{1}{\Phi_-(R)}$ and solve the integral, implying
that
\begin{equation}\label{66}
L =-\Phi_-(R_{0})+\Phi_-(R_{0}+\varepsilon).
\end{equation}

Next, by expanding $V_-(R_{0} + \epsilon)$ into a Taylor series
around $R_{0}$ and focusing only on the first order terms in
$\epsilon$. Because $\epsilon$ is very small, we can ignore the
higher order terms in the expansion. Figure \textbf{4} shows the
behavior of proper length versus thickness. It shows that as the
thickness increases, the proper length of the shell also increases
consistently. This can be attributed to the changing geometric
properties of the shell, where a greater thickness allows for a
larger spatial extent. Understanding this relationship is crucial
for assessing the structural integrity and physical behavior of the
shell, particularly in how it responds to external forces and energy
distributions.
\begin{figure}\center
\epsfig{file=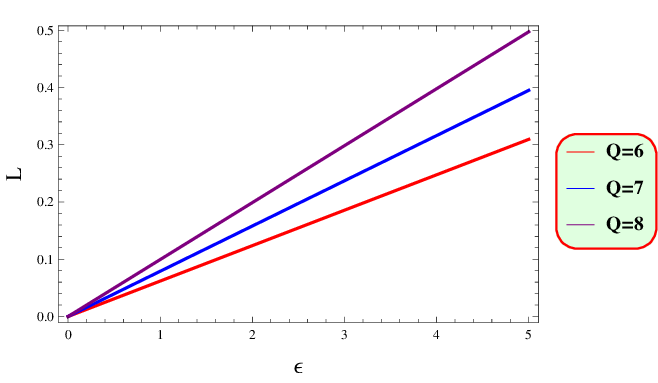,width=.47\linewidth} \caption{Graph of proper
length versus thickness.}
\end{figure}

\subsection{Entropy}

The entropy $S$ for the intermediate thin-shell can be calculated as
\begin{equation}\label{68}
S=\int_{R_{0}}^{R_{0}+\varepsilon}4\pi
R^{2}s(R)\sqrt{e^{\beta(R)}}dR,
\end{equation}
where the entropy density is represented by $s(R)$, and its
mathematical expression is given as follows
\begin{equation}\label{69}
s(R)= \frac{k^{2} (K_{B})^{2}\mathcal{T}(R)}{4 \pi
\hbar^{2}\mathcal{G}}=\frac{k K_{B}}{\hbar}\bigg(\frac{P_{0}}{2 \pi
\mathcal{G}}\bigg)^{\frac{1}{2}},
\end{equation}
here, $k$ is non-zero arbitrary constant, $\mathcal{T}(R)$
represents the temperature, $\hbar$ and $K_{B}$ refer to Planck's
and Boltzmann's constants, respectively. Here, we use the geometric
units, where $G =1$. Consequently, Eq.\eqref{68} simplifies to
\begin{eqnarray}\nonumber
S &=& \int_{R_{0}}^{R_{0}+\varepsilon} 4 \pi R^{2} \frac{k
K_{B}}{\hbar}\bigg(\frac{P_{0}}{2 \pi}\bigg)^{\frac{1}{2}}
\sqrt{e^{\beta(R)}} dR= \int_{R_{0}}^{R_{0}+\varepsilon} 2 \sqrt{2
\pi} R^{2} \frac{k K_{B}}{\hbar} \sqrt{P_{0} e^{\beta(R)}} dR.
\end{eqnarray}
This expression can be further written as
\begin{equation}\label{71a}
S= 2 \sqrt{2 \pi} R^{2} \frac{k K_{B}}{\hbar} N,
\end{equation}
where
\begin{equation}\label{72a}
N = \int_{R_{0}}^{R_{0}+\epsilon}\sqrt{P_{0}
e^{\beta(R)}}dR=\int_{R_{0}}^{R_{0} +\epsilon}D(R)dR.
\end{equation}
According to Eq.\eqref{72a}, calculating the integral directly is
quite difficult at this stage. If we consider $F(R)$ as the integral
of $D(R)$, we can proceed with the calculation. Using this approach,
Eq.\eqref{72a} is simplified after applying the fundamental theorem
of calculus. The result is
\begin{equation}\label{74}
N = [F(R)]_{R_{0}}^{R_{0}+\epsilon} = F(R_{0} + \epsilon) -
F(R_{0}).
\end{equation}
When we keep only the linear terms of $\epsilon$ and expand $F(R_{0}
+ \epsilon)$ using a Taylor series around $d$, we find (using
Eq.\eqref{71a}) that
\begin{eqnarray}\nonumber
S&=&\bigg[\bigg(\frac{\big(2 R_{0}(R_{0}-2 M)+Q^2\big) \big(\zeta
R_{0}^2-2 \psi \big)}{6 Q^2 \psi -4 R_{0}^2 \psi
}-c_{1}\bigg)\bigg(\sqrt{1-\frac{2
M}{R_{0}}+\frac{Q^2}{R_{0}^2}}\\\nonumber&-&\sqrt{\frac{\big(2 R_{0}
(R_{0}-2 M)+Q^2\big) \big(\zeta  R_{0}^2-2 \psi \big)}{6 Q^2 \psi -4
R_{0}^2 \psi }-c_{1}}+\frac{1}{8 \pi}\bigg(\frac{1}{R_{0}^2 }(M
R_{0}-Q^2\\\nonumber&\times&\sqrt{R_{0}^2-2 M R_{0}+Q^2})+\bigg(2
Q^2 \big(9 \zeta M R_{0}^2-6 M\psi -6\zeta R_{0}^3+8 R_{0} \psi
\big)\\\nonumber&-&4 M \big(\zeta R_{0}^4+2 R_{0}^2 \psi \big)-3
\zeta Q^4 R_{0}+4 \zeta R_{0}^5\bigg)\bigg(\sqrt{2}\psi \big(3 Q^2-2
R_{0}^2\big)^2\bigg(\bigg(\big(2
\\\nonumber&\times& c_{1} R_{0} \psi-2 \zeta M R_{0}^2+4 M
\psi +\zeta R_{0}^3-2 R_{0} \psi \big)2 R_{0}+Q^2 \big(\zeta
R_{0}^2-2 (3 c_{1} \psi  \\\label{72}&+&\psi )\big)\bigg)\bigg(\psi
\big(3 Q^2-2
R_{0}^2\big)\bigg)^{-1}\bigg)^{\frac{1}{2}}\bigg)^{-1}\bigg)
\bigg)\bigg]^{\frac{1}{2}}\frac{\big(R_{0}^2 k K_B \epsilon
\big)}{\sqrt{4} \hbar}.
\end{eqnarray}

Figure \textbf{5} shows that the disorder within the gravastar
increases as the shell's thickness expands, with the entropy rising
in proportion to the shell's thickness and peak at the outer
surface. A shell with no thickness results in zero entropy.
Moreover, as the charge increases, the level of disorder also
increases, which helps in forming a more stable structure. The model
parameter affects entropy as well, with positive values leading to
higher entropy than negative ones. This indicates that the outer
layers are more disordered compared to the inner layers. Thus,
increasing charge enhances the disorder within the gravastar,
further promoting structural stability.
\begin{figure}\center
\epsfig{file=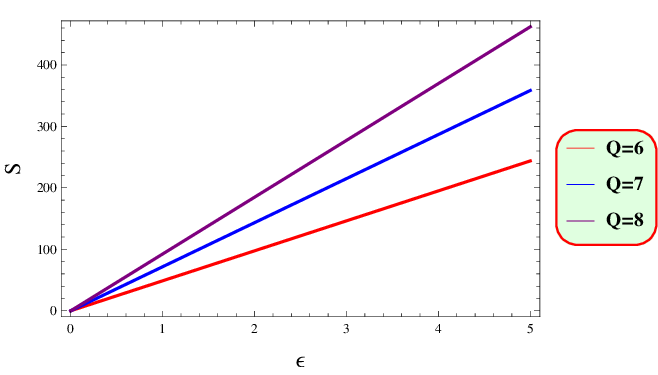,width=.47\linewidth} \caption{Graph of entropy
versus thickness.}
\end{figure}

\subsection{Energy}

We can find the energy of the shell
\begin{equation}\label{76}
E = \int_{R_{0}}^{R_{0}+\varepsilon} 4 \pi R^{2} \rho_{0} dR.
\end{equation}
Using Eqs.\eqref{15} and \eqref{16}, it follows that
\begin{equation}\label{77}
E = \frac{4 \pi  \epsilon ^3}{3}\bigg(\frac{\sqrt{-\frac{2
M}{R_{0}}+\frac{Q^2}{R_{0}^2}+1}-\sqrt{\frac{\big(2 R_{0} (R_{0}-2
M)+Q^2\big) \big(\zeta  R_{0}^2-2 \psi \big)}{6 Q^2 \psi -4 R_{0}^2
\psi }-c_{1}}}{4 \pi R_{0}}\bigg).
\end{equation}
Figure \textbf{6} indicates that the energy of the shell increases
with the shell thickness. In our model, the energy density is
closely related to the pressure within the shell and a thicker shell
allows for greater interactions between gravitational and
electromagnetic forces, which influence the energy distribution.
\begin{figure}\center
\epsfig{file=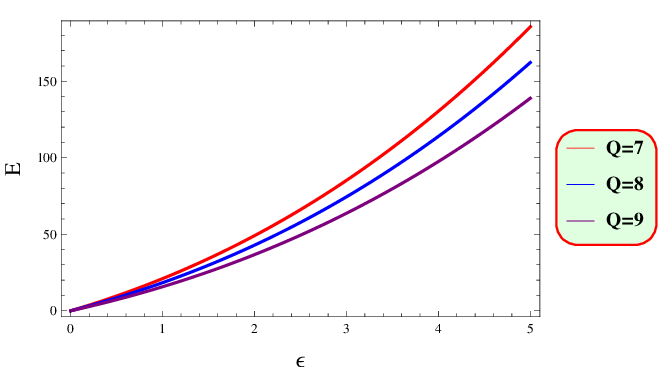,width=.47\linewidth} \caption{Graph of energy
versus thickness.}
\end{figure}

\subsection{The EoS Parameter}

The EoS parameter can be expressed as
\begin{eqnarray}\nonumber
\omega&=& \bigg[\bigg(\sqrt{1-\frac{2
M}{R_{0}}+\frac{Q^2}{R_{0}^2}}-\sqrt{\frac{\big(2 R_{0} (R_{0}-2
M)+Q^2\big) \big(\zeta  R_{0}^2-2 \psi \big)}{6 Q^2 \psi -4 R_{0}^2
\psi }-c_{1}}\\\nonumber&+&\frac{1}{8 \pi}\bigg(\frac{1}{R_{0}^2 }(M
R_{0}-Q^2\sqrt{R_{0}^2-2 M R_{0}+Q^2})+\bigg(2 Q^2 \big(9 \zeta M
R_{0}^2-6 M
\\\nonumber&\times&\psi -6\zeta R_{0}^3+8 R_{0} \psi \big)-4 M \big(\zeta
R_{0}^4+2 R_{0}^2 \psi \big)-3 \zeta Q^4 R_{0}+4 \zeta
R_{0}^5\bigg)\bigg(\sqrt{2}
\\\nonumber&\times&\psi \big(3 Q^2-2
R_{0}^2\big)^2 \bigg(\bigg(\big(2 c R_{0} \psi-2 \zeta M R_{0}^2+4 M
\psi +\zeta R_{0}^3-2 R_{0} \psi \big)2 R_{0} \\\nonumber&+&Q^2
\big(\zeta R_{0}^2-2 (3 c_{1} \psi +\psi )\big)\bigg)\bigg(\psi
\big(3 Q^2-2
R_{0}^2\big)\bigg)^{-1}\bigg)^{\frac{1}{2}}\bigg)^{-1}\bigg)
\bigg)\bigg(\frac{1}{4 \pi
R_{0}}\\\nonumber&\times&\bigg(\sqrt{1-\frac{2
M}{R_{0}}+\frac{Q^2}{R_{0}^2}}-\sqrt{\frac{\big(2 R_{0} (R_{0}-2
M)+Q^2\big) \big(\zeta  R_{0}^2-2 \psi \big)}{6 Q^2 \psi -4 R_{0}^2
\psi }-c_{1}}\bigg)\bigg)^{-1}\bigg].
\end{eqnarray}
The validity of different stellar models is effectively
characterized by various EoS parameter values. For instance,
$\omega=0$ corresponds to flat surfaces with non-relativistic fluid,
$\omega= -\frac{1}{3}$ applies to curved surfaces, $\omega= -1$ is
associated with DE, $\omega< -1$ signifies phantom energy and
$\omega \geq -1$ defines the non-phantom regime. In the Figure
\textbf{7}, we observe that $\omega$ drops below -1, entering the
phantom region, which shows the significance of the gravastar
structure. This behavior implies a negative energy density
associated with the gravastar, suggesting the possibility of
repulsive gravitational effects that could contribute to the
accelerated expansion of the universe. In contrast to conventional
matter and energy, which have $\omega \geq -1$, gravastars' ability
to enter the phantom regime positions them as viable alternative
models for explaining cosmic phenomena such as DE, that standard
cosmological models struggle to address.
\begin{figure}\center
\epsfig{file=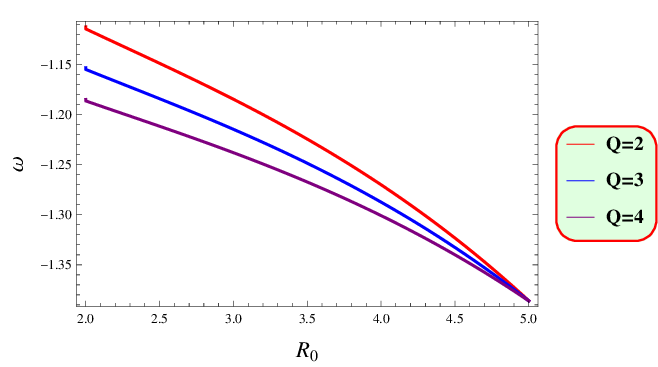,width=.47\linewidth} \caption{Graph of EoS
versus $R_{0}$.}
\end{figure}

\subsection{Stability Analysis}

Here, we assess the stability of thin-shell gravastars by analyzing
the potential function and its second derivative. A structure is
considered stable when $ V^{\prime\prime}(R_{0}) > 0 $, unstable
when $V^{\prime\prime}(R_{0}) < 0$, and unpredictable for
$V^{\prime\prime}(R_{0}) = 0$. We give the graphical behavior of
$V^{\prime\prime}(R_{0})$ for all the three black holes due to
complicated expressions. Figures \textbf{8-10} confirm that
$V^{\prime\prime}(R_{0})>0$ for three distinct values of $Q,~e$ and
$l$, hence the stable configuration of thin-shell gravastars is
achieved. A positive second derivative signifies that small
perturbations in the configuration will generate restoring forces,
returning the system to its equilibrium state. Thus, the gravastar
configuration is stable against minor changes, whether caused by
external influences or internal fluctuations.
\begin{figure}\center
\epsfig{file=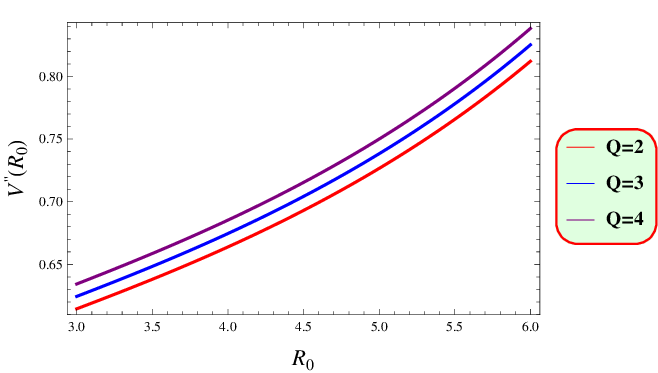,width=.47\linewidth} \caption{Potential for the
shell beyond RN metric.}
\end{figure}
\begin{figure}\center
\epsfig{file=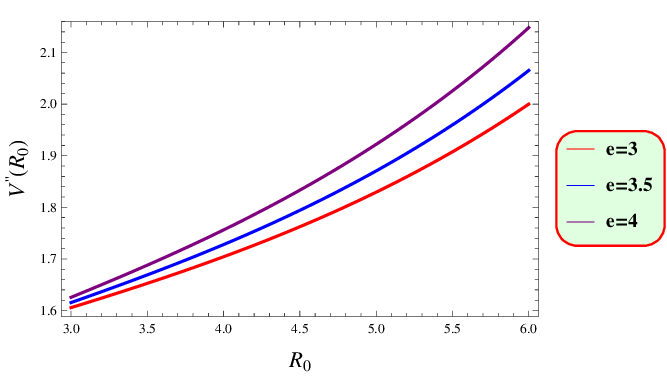,width=.47\linewidth} \caption{Potential for the
shell beyond Bardeen black hole.}
\end{figure}
\begin{figure}\center
\epsfig{file=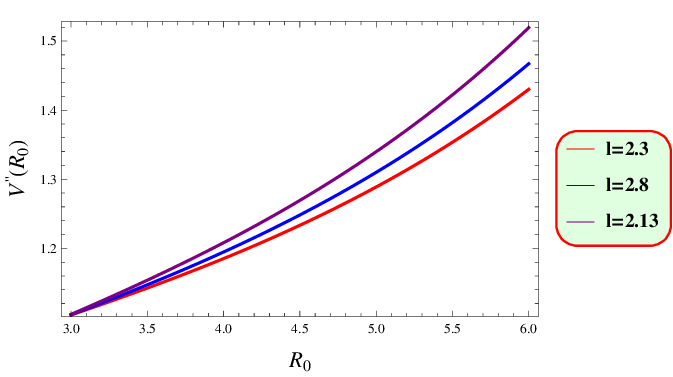,width=.47\linewidth} \caption{Potential for the
shell beyond Hayward black hole.}
\end{figure}

Next, we discuss the stability analysis through causality condition
and adiabatic index. In general fluids, the speed of sound, which
measures how quickly disturbances move through the medium, can be
expressed as
\begin{equation}\label{80}
v_{s}^{2} = \frac{p^{\prime}}{\varrho^{\prime}},
\end{equation}
where $\varrho$ and $p$ for the thin-shell are given in
Eqs.\eqref{15} and \eqref{14}, respectively. To find the speed of
sound in thin-shell, this expression can be evaluated at $R$.
Poisson and Visser \cite{4a} proposed that, due to causality, the
speed of sound should not surpass the speed of light (with $c= 1$ in
natural units). Thus, the speed of sound $v_{s}^{2}$ is expected to
fall within the range $ 0 \leq v_{s}^{2} \leq 1$.
\begin{figure}\center
\epsfig{file=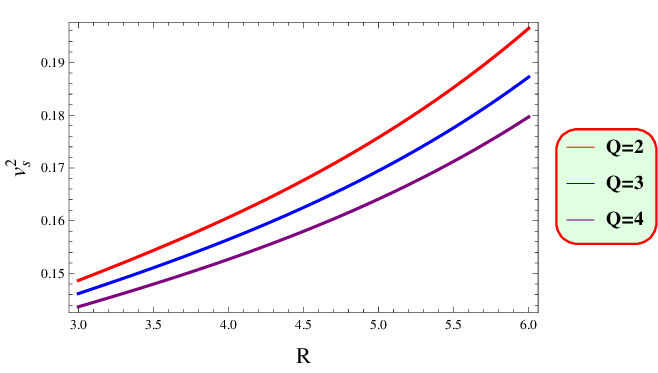,width=.47\linewidth} \caption{Graph of
$v_{s}^{2}$ with respect to $R$ for different values of $Q$.}
\end{figure}

The stability of gravastars has been extensively analyzed using this
criterion. For instance, Lobo \cite{4b} applied this approach to
determine stability limits for a wormhole in the presence of a
cosmological constant. Other studies \cite{lm} have also
investigated the stability of charged gravastars. However, as
mentioned in \cite{4a}, there are some limitations of using the
speed of sound for stability analysis. In the stiff matter region
(where $\omega = 1$), it is not entirely clear if Eq.\eqref{80}
accurately represents the speed of sound. This uncertainty arises
from the incomplete understanding of the microscopic properties of
stiff matter, making the typical fluid argument potentially
unreliable. Despite not being a complete condition, this method
still provides a necessary condition for ensuring the stability of
the thin-shell surrounding the gravastar. Inserting the required
values in \eqref{80}, we derive the result for $v_{s}^{2}$.

Figure \textbf{11} illustrates how the changes in $v_{s}^{2}$, in
relation to the radial distance $R$ and different charge values,
influence the stability of the gravastar's shell. The fact that the
condition $ 0 \leq v_{s}^{2} \leq 1$ is satisfied indicates that the
gravastar maintains stability under these conditions. Our analysis
indicates that increasing charge affects both energy density and
pressure distribution, subsequently influencing the $v_{s}^{2}$.
This insight enhances our understanding of the gravastar's physical
behavior.

The adiabatic index $(\Gamma)$ is another factor for checking the
stability of stars, including gravastars. It shows how pressure and
density affect a star's stability and is calculated using a specific
formula
\begin{equation}\nonumber
\Gamma=\frac{v_{s}^{2} (p+\varrho)}{p}.
\end{equation}
To analyze stability, it is important to find the value of $\Gamma$.
For an object to be stable, $\Gamma$ should be greater than 4/3. If
$\Gamma$ drops below this value, the object becomes unstable and can
collapse \cite{52a}.
\begin{figure}\center
\epsfig{file=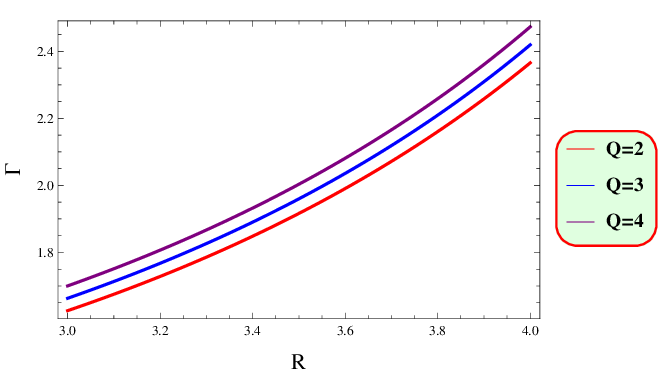,width=.47\linewidth} \caption{Variation of
$\Gamma$ with respect to $R$ for different values of $Q$.}
\end{figure}
Figure \textbf{12} demonstrates that our system remains stable, as
it fulfills the required limit ($\Gamma>\frac{4}{3}$). Thus, we
achieve a viable and stable gravastar. This condition ensures that
the gravastar can maintain its structure against gravitational
collapse and energy dissipation. Our analysis emphasizes key
parameters influencing stability such as energy density and
pressure, along with their interrelationships.

\section{Summary and Discussion}

In this paper, we have investigated a charged gravastar solution
using the Finch-Skea metric within the framework of $f(\mathbb{Q})$
gravity. To gain a deeper understanding of the gravastar, we have
examined three distinct regions: the interior core, the intermediate
shell and the exterior vacuum, each governed by its own specific EoS
parameter. We have analyzed key physical properties, including
proper length, entropy, energy and the EoS parameter. The stability
of the gravastar solution has been assessed through the effective
potential, the causality condition and adiabatic index. We have
identified several key aspects of the gravastar solutions, which are
summarized as follows.
\begin{itemize}
\item We have found that both the increasing energy density and pressure
indicate that the shell's outer boundary is denser than its inner
boundary (Figures \textbf{1-3}).
\item The proper length steadily increases with the shell's thickness
for different charge values. This illustrates a fundamental physical
characteristic of the shell's structure (Figure \textbf{4}).
\item We have found that the disorder within the gravastar increases
as the shell's thickness expands. Furthermore, higher values of
charge and positive model parameters lead to greater entropy,
contributing to a more stable structure (Figure \textbf{5}).
\item We have observed that the shell energy steadily increases with thickness,
hence the gravastar remains in stable position (Figure \textbf{6}).
\item The EoS is found to drop below -1, moving
into the phantom region for various charge values (Figure
\textbf{7}).
\item The stable configuration of thin-shell
gravastars has been checked using the condition
$V^{\prime\prime}(R_{0})>0$, (Figures \textbf{8-10}). We have also
assessed the stability of the gravastar through speed of sound and
adiabatic index and have found that the gravastar satisfies the
stability criteria (Figure \textbf{11,12}).
\end{itemize}

Gravastars have garnered significant attention in modified gravity
theories due to their potential to offer alternative explanations
for compact stellar structures. This study focuses on gravastars
within the $f(\mathbb{Q})$ gravity framework, a theory renowned for
its consistency with observational data and its capability to
explain the universe accelerated expansion. Unlike GR, which relies
heavily on DE, $f(\mathbb{Q})$ gravity provides a geometric
foundation to address challenges such as cosmic acceleration and
singularities. In this paper, gravastar models offer a distinct
advantage by eliminating event horizons, thereby resolving the
information loss paradox. These configurations are characterized by
finite pressure and density, presenting them as stable and
physically consistent alternatives to black holes. Furthermore,
$f(\mathbb{Q})$ gravity enables a deeper exploration of
gravitational and electromagnetic interactions, contributing to the
broader understanding of compact astrophysical phenomena.

We have identified a set of viable solutions that successfully avoid
the central singularities and event horizons typically associated
with black holes. The presence of charge introduces an
outward-directed force, providing a repulsive effect that prevents
the gravastar from collapsing into a singularity. As a result,
charged gravastar demonstrate greater stability compared to
uncharged gravastar. Future research could delve further into
gravastar solutions within the context of $f(\mathbb{Q})$ gravity.
Our findings reveal a consistent enhancement of physical properties
due to the influence of charge, in line with observations in GR and
other modified gravity theories \cite{4e}-\cite{4d}.

Building on this, we compare our findings with existing studies to
highlight the novel contributions of our work. In a recent paper
\cite{8aa}, Mohanty and Sahoo used the Krori-Barua metric in
$f(\mathbb{Q})$ gravity to explore gravastar characteristics, such
as proper length, entropy and stability condition. Our analysis
emphasizes significant role of charge in enhancing gravastar
stability. We show that higher charge values increase entropy and
stability, providing a more understanding of gravastar behavior as
compared to Mohanty and Sahoo. Shamir and Ahmad \cite{m2} examined
gravastars in $f(\mathbb{G}, T)$ gravity and found that their study
does not emphasize the repulsive forces introduced by charge, a
critical aspect in our study. Yousaf et al. \cite{2sd} explored
charged gravastars in $f(\mathbb{R}, T)$ gravity, focusing on the
role of electromagnetic fields in gravastar stability. Sharif and
Naz \cite{1gg} studied gravastars in $f(\mathbb{R}, T^{2})$ gravity
using the Finch-Skea metric. Our work not only aligns with these
findings but also extend them by providing a clearer correlation
between charge and enhanced stability in the $f(\mathbb{Q})$
context.

The treatment of the Finch-Skea metric significantly influences the
results by simplifying field equations and ensuring singularity-free
solutions. It maintains regularity at the center, smooth matching
between regions, and stability of energy density and pressure
profiles, even with the inclusion of charge. Through Israel junction
conditions, surface energy density and pressure are accurately
calculated, ensuring physical consistency. Its compatibility with
exterior geometries enables unified solutions for charged
configurations. In this work, we have analyzed key physical
quantities such as proper length, entropy, energy density, pressure,
and EoS parameter, specifically at equilibrium, along with stability
indicators like the causality condition and adiabatic index, within
this gravity. While these analyses align with established
theoretical approaches, they offer new insights into the stability
and equilibrium of the gravastar model. Our graphical results
further confirm that the proposed model is both stable and
theoretically viable.\\\\
\textbf{Data Availability Statement:} No data was used for the
research described in this paper.

\end{document}